\documentclass{JHEP3}

\title{Type I Supergravity Effective Action from Pure Spinor Formalism}
\author{Geov\'a Alencar$^a$ \\
$^a$Instituto de f\'{\i}sice te\'{o}rica, Universidade Estadual Paulista\\
Rua Pamplona 145, 01405-900, S\~{a}o Paulo, SP, Brasil}
\abstract {Using the pure spinor formalism, we compute the tree-level
correlation functions for three strings, one closed and two open, in
$N=1$ $D=10$ superspace. Expanding the superfields in components, the
respective terms of the effective action for the type I supergravity
are obtained. All terms found agree with the effective action
known in the literature. This result gives one more consistency test
for the pure spinor formalism.}

\begin{document}
\section{Introduction}

The covariant quantization of superstring theory has been an unresolved
problem for a long time. The covariant quantization, besides having manifest
supersymmetry, makes the computation of scattering amplitudes easier. This
is important for understanding the low energy limit of superstrings, through
the construction of effective actions corresponding to such amplitudes. In
order to solve the problem of manifest covariant quantization, a new
formalism, known as pure spinors formalism, was proposed \cite{Nathan 2000}.
This new formalism keeps all the good properties of Ramond-Neveu-Schwarz and
Green-Schwarz and does not have its undesired characteristics. In the
Ramond-Neveu-Schwarz formalism, when the number of loops in computations of
scattering amplitudes are increased, more and more spin structures have to
be considered which makes the computations very long. On the other hand, in
the Green-Schwarz formalism the quantization is only possible in the
light-cone gauge and the amplitude computations involve non-covariant
operators at the interaction points.

The complete equivalence between the pure spinor formalism and other
formalisms is missing. Until this point is reached, the formalism needs to
pass many consistency tests. One of these tests consists of computing
scattering amplitudes and comparing the results with those coming from other
formalisms. These tests have been carried out for amplitudes involving
closed superstrings at one loop \cite{mafra 1 loop} and two loops \cite%
{mafra 2 loops}, among others. Interestingly, amplitudes involving mixed
superstrings have not been considered in the literature. One important point
is the fact that in pure spinor formalism the amplitudes have explicit
super-Poincar\'e symmetry, making the results automatically supersymmetric
since the beginning.

From the viewpoint of field theory, the effective action for the type I
supergravity is obtained from the global super Yang-Mills action by imposing
local supersymmetry, which originates many compensation terms \cite%
{bergshoeff} that are interpreted as interaction terms. From the viewpoint
of superstrings, all these interaction terms must come out naturally from
amplitude computations. The interaction terms of the type I supergravity
effective action have some interesting properties. A very peculiar one is
the fact that there is a coupling between the Kalb-Ramond and two photons.
This term is needed to obtain local supersymmetry. In order to keep gauge
invariance, the Kalb-Ramond field must have a unusual transformation under
U(1) symmetry. This coupling will become very important for the mixed
anomaly cancelation in the SO(32) theory.

As pointed above, all these terms must come naturally from superstring
theory. However, as they involve gravitational and Yang-Mills fields,
amplitudes with open and closed strings have to be considered. In this work,
we will show that these terms can be obtained from the pure spinor
formalism. In string theory, the effective action can be obtained
considering the scattering amplitudes or correlation functions of three
points given by

\[
\mathcal{A}=\langle V_{1}V_{2}V_{3}\rangle,
\]
where the Vs above represent the physical states and are called vertex
operators. For the mixed scatterings, we must use the upper-half complex
plane. The closed string is represented by a point in the interior of the
plane, while the open strings, by points in the real axis. The number of
conformal Killing vectors in this case is three, and the number of moduli is
zero. This makes it possible to fix the positions of the closed string and
of one of the open strings, obtaining

\[
\mathcal{A}=\langle V_{1}V_{2}\int U_{3}\rangle.
\]

In the last equation, $V_{1}$ represents the fixed closed string, $V_{2}$
represents the fixed open string and $U_{3}$, the integrated open string.

In the second section of this paper, we give the expression for the type I
supergravity effective action and its respective linearized action. This
will be compared with the expression obtained from the pure spinor
formalism. In the third section, the prescription for amplitudes in pure
spinor formalism will be briefly summarized, and an expression for the
computation of one closed and two open strings will be given. Using
superspace identities, a simple expression will be found following identical
steps of \cite{mafra identities}. This expression is given by

\[
\mathcal{A}=g_{o}^{\prime 2}g_{c}^{\prime }\pi i\alpha ^{\prime
}\left\langle \left[ A_{m}^{1}\left( \lambda \tilde{A}^{1}\right) +\tilde{A}%
_{m}^{1}\left( \lambda A^{1}\right) \right] \left( \lambda A^{2}\right)
\left( \lambda \gamma ^{m}W^{3}\right) \right\rangle ,
\]
where the superfields in the closed string vertex operator $V^1$
have been written as the product of two open string superfields.

It is important to note that this expression is manifestly super-Poincar\'e
covariant and that all the amplitudes involving one closed and two open
massless strings are contained in it. This shows one of the advantages of
the pure spinor formalism. It will be also shown, in appendix B, that this
expression has all the gauge invariances.

In the fourth section, with this simple expression, the superfields will be
expanded in components, and explicit results will be given for all the
correlation functions from which the effective action can be obtained. All
cases will be considered with some details, except the case of one
gravitino, one photon and one photino, which will be left for appendix D.

In appendix A, some useful identities will be given, and in appendix C,
the correlation function for graviton-photon-photon will be considered
in the Ramond-Neveu-Schwarz formalism for matter of comparison with
pure spinor formalism.

\section{Type I supergravity effective action}

As said before, the Type I effective action can be obtained by imposing
local supersymmetry in the Super-Maxwell action and is given by\cite%
{bergshoeff}

\begin{eqnarray*}
\frac{1}{\sqrt{-G}} &&\mathcal{L=}-\frac{1}{2\kappa ^{\prime }}\tilde{R}-
\frac{9}{16\kappa^{\prime 2}}\left( \frac{\partial _{m}\phi }{\phi }\right) ^{2}-
\frac{1}{4}\phi ^{-3/4}\left( F_{mn}\right) ^{2}-\frac{3}{4}\phi
^{-3/2}H_{mnp}^{2} \\
&&-\frac{1}{2}\psi _{m}\gamma ^{mnp}D_{n}\psi _{p}-\frac{1}{2}\lambda \gamma
^{m}D_{m}\lambda -\frac{1}{2}\chi \gamma ^{m}D_{m}\chi \\
&&-\frac{1}{4}\kappa ^{\prime }\phi ^{-3/8}\chi \gamma ^{m}\gamma
^{np}F_{np}\left( \psi _{m}+\frac{\sqrt{2}}{12}\gamma _{m}\lambda \right) \\
&&+\frac{\sqrt{2}}{16}\kappa ^{\prime }\phi ^{-3/4}\chi \gamma ^{mnp}\chi
H_{mnp}-\frac{3\sqrt{2}}{8}\psi _{m}\gamma ^{n}\gamma ^{m}\lambda \left(
\frac{\partial _{n}\phi }{\phi }\right) \\
&&+\frac{\sqrt{2}\kappa ^{\prime }}{16}\phi ^{-3/4}H_{npq}(\psi _{m}\gamma
^{mnpqr}\psi _{r}+6\psi ^{n}\gamma ^{p}\psi ^{q} \\
&&-\sqrt{2}\psi _{m}\gamma ^{npq}\gamma ^{m}\lambda )+\left( Fermi\right)
^{4}.
\end{eqnarray*}

In order to go to the String Frame, we make the field redefinitions
\begin{eqnarray*}
\tilde{G}_{mn} &=&e^{2\omega }G_{mn},\phi =e^{8\omega /3},\psi _{m}=\frac{1}{%
\kappa ^{\prime }}e^{\omega /2}\psi _{m}^{\prime },\chi =\frac{1}{g^{\prime }%
}e^{-5\omega /2}\xi \\
\lambda &=&\frac{1}{\kappa ^{\prime }}e^{-\omega /2}\lambda ^{\prime },A_{m}=%
\frac{1}{g^{\prime }}A_{m}^{\prime },B_{mn}=\frac{\kappa ^{\prime }}{3\sqrt{2%
}g^{\prime 2}}B_{mn}^{\prime },\eta =e^{\omega /2}\eta ^{\prime },
\end{eqnarray*}
where $\omega =( \Phi _{0}-\Phi )/4$. We obtain the following Lagrangian

\begin{eqnarray*}
&&\mathcal{L}=\frac{\sqrt{-G}}{2\kappa ^{2}}e^{-2\Phi }\left[ R+4\partial
_{m}\Phi \partial ^{m}\Phi \right] -\frac{\sqrt{-G}}{4\kappa^{2}}e^{-\Phi
}F_{mo}^{2}-\frac{1}{24\kappa ^{2}}\sqrt{-G}H_{mnp}^{\prime 2} \\
&&-\frac{1}{2\kappa ^{2}}\sqrt{-G}e^{-2\Phi }\psi _{m}\gamma ^{mnp}D_{n}\psi
_{p}-\frac{1}{\kappa ^{2}}\sqrt{-G}e^{-2\Phi }\lambda \gamma ^{m}D_{m}\lambda
\\
&&-\frac{1}{2\kappa ^{2}}\sqrt{-G}e^{-\Phi }\xi \gamma ^{m}D_{m}\xi -\frac{1%
}{4\kappa ^{2}}\sqrt{-G}e^{-\Phi }\xi \gamma ^{m}\gamma ^{np}F_{np}\left(
\psi _{m}+\frac{1}{6}\gamma _{m}\lambda \right) \\
&&+\frac{1}{48}\frac{1}{\kappa ^{2}}\sqrt{-G}\xi \gamma ^{mnp}\xi
H_{mnp}^{\prime }-\frac{2}{\kappa^{2}}\sqrt{-G}e^{-2\Phi }\psi _{m}\gamma
^{n}\gamma ^{m}\lambda \partial _{n}\Phi \\
&&+\frac{3}{8\kappa ^{2}}\sqrt{-G}e^{-\Phi }H_{npq}^{\prime }(\psi
_{m}\gamma ^{mnpqr}\psi _{r}+6\psi ^{n}\gamma ^{p}\psi ^{q}-2\psi _{m}\gamma
^{npq}\gamma ^{m}\lambda ).
\end{eqnarray*}

In the last equation, $\psi $ is the gravitino, $\lambda$ is the dilatino, $%
\xi$ is the photino and we use the standard notation for the bosonic fields.
The coupling between the Kalb-Ramond field and the photon comes from $%
H^{\prime mnp}$, which is defined as

\[
H^{\prime }=dB+3AdA.
\]

In the low energy limit, we need only the linearized Lagrangian. The usual
procedure is to make
\[
G^{mn}=\eta ^{mn}+h^{mn}.
\]

In order to simplify some terms, we use the identities
\[
\gamma ^{m}\gamma ^{no}-\gamma ^{no}\gamma ^{m}=-2G^{mo}\gamma
^{n}+2G^{mn}\gamma ^{o}
\]
and
\[
\gamma ^{m}\gamma _{np}\gamma _{m}=6\gamma _{np}.
\]

Regarding only the terms related to the two open and one closed string
amplitudes, we obtain
\begin{eqnarray}
&&\mathcal{L}=-\frac{1}{2\kappa^{2}}h_{n}^{m}F_{mo}F^{no}-\frac{\Phi }{4\kappa^{2}}
F_{mo}^{2}-\frac{1}{8\kappa ^{2}}H_{mnp}A^{m}F^{np}  \nonumber \\
&&-\frac{1}{2\kappa^2}h_{mn}\xi \gamma ^{m}\partial ^{n}\xi +\frac{1}{48}\frac{1
}{\kappa ^{2}}\xi \gamma ^{mnp}\xi H_{mnp}
\label{Acao efetiva supergravidade} \\
&&-\frac{1}{\kappa ^{2}}\xi \gamma _{p}\psi _{m}F^{mp}-\frac{1}{4\kappa ^{2}}
\xi \gamma ^{np}\lambda F_{np}.  \nonumber
\end{eqnarray}

The indices here are raised with $\eta $. Attention is required in
considering the dilaton contribution coming from the expansion of the metric
determinant, because
\[
\sqrt{-G}=1+\frac{1}{2}h_{m}^{m}
\]
and the trace of $h_{mn}$ is related to the dilaton.

\section{The tree-level correlation function for one closed and two open
massless strings in the pure spinor formalism}

As discussed in the introduction, from the viewpoint of string theory, the
graviton is represented by a closed string and the photon by an open string.
In pure spinor formalism, the fixed operator for the open string is given by
\cite{Nathan 2000}

\[
V=g_{o}^{\prime }\lambda ^{\alpha }A_{\alpha },
\]
where $\lambda $ is a pure spinor satisfying

\begin{equation}
\lambda \gamma ^{m}\lambda =0 .  \label{vinculo}
\end{equation}

The BRST operator is given by

\[
Q=\lambda ^{\alpha }d_{\alpha }
\]
with

\[
d_{\alpha }=\frac{\alpha ^{\prime }}{2}p_{\alpha }-\frac{1}{2}\theta \gamma
^{m}\partial{x}_{m}-\frac{1}{8}\gamma _{\alpha \beta }^{m}\gamma _{m\delta \eta
}\theta ^{\beta }\theta ^{\delta }\partial \theta ^{\eta }.
\]

The physical state condition gives us the equations of motion

\begin{equation}
D_{\alpha }A_{\beta }+D_{\beta }A_{\alpha }=\gamma _{\alpha \beta }^{m}A_{m}
,  \label{Eq of motion}
\end{equation}
where

\[
D_{\alpha }=\frac{\alpha ^{\prime }}{2}\partial _{\alpha }+\theta \gamma
^{m}\partial _{m} , \ \ \bar{D}_{\alpha }=
\frac{\alpha ^{\prime }}{2} \bar{\partial} _{\alpha }+\bar{\theta} \gamma
^{m} \partial _{m}\]
and $A_{m}$ is a vector superfield. The integrated vertex operator for the
open string is

\[
g_{o}^{\prime }\int dy_{3}\left( \partial \theta ^{\alpha }A_{\alpha
}+A_{m}\Pi ^{m}+d_{\alpha }W^{\alpha }+\frac{1}{2}N^{nm}\mathcal{F}%
_{nm}\right) ,
\]
where $A_{\alpha }$, $A_{m}$ and $d_{\alpha }$ are defined above and $%
W^{\alpha }$ and $F_{mn}$ are field strengths given by

\[
W^{\alpha }=\frac{1}{10}\gamma _{m}^{\alpha \beta }D_{\beta }A^{m}, \ \
\mathcal{F}_{mn}=2\partial _{\lbrack m}A_{n]},
\]
being $N^{mn}$ the Lorentz generators for the ghosts $\lambda $, given by

\[
N^{nm}=\frac{\alpha ^{\prime }\left( \lambda \gamma ^{mn}\omega \right) }{4}
.
\]

When necessary, the superfields will be expanded in components.The vertex 
operator for the closed string is given by the product of two open string
operators $\left( \lambda ^{\alpha }A_{\alpha }\right) \left( \bar{\lambda}%
^{\alpha }\bar{A}_{\alpha }\right) $. Then, we have for the amplitude\cite%
{Nathan 2000}

\begin{eqnarray}
\mathcal{A} &=&\left\langle V_{g}V_{h}\int U_{h}\right\rangle =g_{o}^{\prime
2}g_{c}^{\prime }\int dy_{3}\left\langle \left[ \left( \lambda ^{\alpha
}A_{\alpha }^{1}\left( z\right) \right) \left( \bar{\lambda}^{\alpha }\bar{A}%
_{\alpha }^{1}\left( \bar{z}\right) \right) \right] \left( \lambda
A^{2}\left( y_{2}\right) \right) U_{h}\left( y_{3}\right) \right\rangle
\nonumber \\
&=&g_{o}^{\prime 2}g_{c}^{\prime }\int dy_{3}\langle \left[ \left( \lambda
^{\alpha }A_{\alpha }^{1}\left( z\right) \right) \left( \bar{\lambda}%
^{\alpha }\bar{A}_{\alpha }^{1}\left( \bar{z}\right) \right) \right] \left(
\lambda A^{2}\left( y_{2}\right) \right) \times  \label{amplitude inicial ps}
\\
&&\left( \partial \theta ^{\alpha }A_{\alpha }^{3}+A_{m}^{3}\Pi
^{m}+d_{\alpha }W_{3}^{\alpha }+\frac{1}{2}N^{nm}\mathcal{F}_{nm}^{3}\right)
\rangle \nonumber .
\end{eqnarray}

A way to simplify an expression similar to this is given in (\cite{mafra
identities}). We must show here that the same expression is valid in the
case of mixed string amplitudes following the same steps. First of all we
must note that the first term of the integrated operator has null OPE with
the other vertex operators. After this, we get

\begin{eqnarray}
\mathcal{A} &=&g_{o}^{\prime 2}g_{c}^{\prime }\int dy_{3}\langle \left[
\left( \lambda ^{\alpha }A_{\alpha }^{1}\left( z\right) \right) \left( \bar{%
\lambda}^{\alpha }\bar{A}_{\alpha }^{1}\left( \bar{z}\right) \right) \right]
\left( \lambda A^{2}\left( y_{2}\right) \right) \times
\label{amplitude ps geral} \\
&&\left( A_{m}^{3}\Pi ^{m}+d_{\alpha }W_{3}^{\alpha }+\frac{1}{2}N^{nm}%
\mathcal{F}_{nm}^{3}\right) \rangle \nonumber .
\end{eqnarray}

We also have that the OPE of any vertex operator with $A^{2}\left(
y_{2}\right) $ will have null result by fixing $y_{2}\rightharpoonup \infty $
. For the next term, we need the standard OPE

\[
:\Pi ^{m}\left( y_{3}\right) e^{ik\cdot x}\left( z\right): \sim
ik_{1}^{m}\alpha \prime \left[ \frac{1}{z-y_{3}}+\frac{1}{\bar{z}-y_{3}}%
\right] e^{ik\cdot x}\left( z\right)
\]
to obtain

\begin{equation}
i\alpha ^{\prime }\int dy_{3}k_{1}^{m}\alpha \prime \left[ \frac{1}{z-y_{3}}+%
\frac{1}{\bar{z}-y_{3}}\right] \left\langle \left[ \left( \lambda ^{\alpha
}A_{\alpha }^{1}\left( z\right) \right) \left( \bar{\lambda}^{\alpha }\bar{A}%
_{\alpha }^{1}\left( \bar{z}\right) \right) \right] \left( \lambda
A^{2}\left( y_{2}\right) \right) A_{m}^{3}\right\rangle \label{segundo
termo ps}.
\end{equation}

Now fixing
\[
\mbox{Im}(z)=ia,\ \ \mbox{Re}(z)=0
\]
the term in eq. (\ref{segundo termo ps}) is also null

\[
i\alpha ^{\prime }\int dy_{3}k_{1}^{m}\alpha \prime \left[ \frac{1}{ia-y_{3}}+%
\frac{1}{-ia-y_{3}}\right] \left\langle \left[ \left( \lambda ^{\alpha
}A_{\alpha }^{1}\left( z\right) \right) \left( \bar{\lambda}^{\alpha }\bar{A}%
_{\alpha }^{1}\left( \bar{z}\right) \right) \right] \left( \lambda
A^{2}\left( y_{2}\right) \right) A_{m}^{3}\right\rangle =0,
\]%
where in the above expression a contour integral gives a null result. Them only the two last terms contribute

\[
\mathcal{A}=g_{o}^{\prime 2}g_{c}^{\prime }\int dy_{3}\left\langle \left[
\left( \lambda ^{\alpha }A_{\alpha }^{1}\left( z\right) \right) \left( \bar{%
\lambda}^{\alpha }\bar{A}_{\alpha }^{1}\left( \bar{z}\right) \right) \right]
\left( \lambda A^{2}\left( y_{2}\right) \right) \left( d_{\alpha
}W_{3}^{\alpha }+\frac{1}{2}N^{nm}\mathcal{F}_{nm}^{3}\right) \right\rangle
.
\]

For the next term, we must use the OPE

\[
d_{\alpha }\left( z_{i}\right) V\left( z_{j}\right) \sim -\frac{\alpha
^{\prime }}{2}\frac{D_{\alpha }V}{z_{j}-z_{i}}-\frac{\alpha ^{\prime }}{2}%
\frac{\bar{D}_{\alpha }V}{\bar{z}_{j}-z_{i}}
\]
and we arrive in
\begin{eqnarray*}
\mathcal{A}_{1} &=&g_{o}^{\prime 2}g_{c}^{\prime }\frac{\alpha ^{\prime }}{2}%
\int \frac{dy_{3}}{z-y_{3}}\left\langle D_{\alpha }\left( \lambda
A^{1}\left( z\right) \right) \left( \overline{\lambda A}^{1}\left( \bar{z}%
\right) \right) \left( \lambda A^{2}\right) W_{3}^{\alpha }\right\rangle \\
&&-g_{o}^{\prime 2}g_{c}^{\prime }\frac{\alpha ^{\prime }}{2}\int \frac{%
dy_{3}}{\bar{z}-y_{3}}\left\langle \bar{D}_{\alpha }\left( \lambda
A^{1}\left( z\right) \right) \left( \overline{\lambda A}^{1}\left( \bar{z}%
\right) \right) \left( \lambda A^{2}\right) W_{3}^{\alpha }\right\rangle .
\end{eqnarray*}

Fixing $z=ia$ in the last equation and solving the integrals we get

\[
\mathcal{A}_{1}=+g_{o}^{\prime 2}g_{c}^{\prime }\pi i\alpha ^{\prime
}\left\langle D_{\alpha }\left( \lambda A^{1}\right) \left( \overline{%
\lambda A}^{1}\right) \left( \lambda A^{2}\right) W_{3}^{\alpha
}\right\rangle +g_{o}^{\prime 2}g_{c}^{\prime }\pi i\alpha ^{\prime
}\left\langle \left( \lambda A^{1}\right) \bar{D}_{\alpha }\left( \overline{%
\lambda A}^{1}\right) \left( \lambda A^{2}\right) W_{3}^{\alpha
}\right\rangle .
\]

After solving the OPEs, we just have zero modes and there is no difference
between holomorphic and antiholomorphic terms, i.e.,

\[
\mathcal{A}_{1}=+g_{o}^{\prime 2}g_{c}^{\prime }\pi i\alpha ^{\prime
}\left\langle D_{\alpha }\left( \lambda A^{1}\right) \left( \lambda \tilde{A}%
^{1}\right) \left( \lambda A^{2}\right) W_{3}^{\alpha }\right\rangle
+g_{o}^{\prime 2}g_{c}^{\prime }\pi i\alpha ^{\prime }\left\langle \left(
\lambda A^{1}\right) D_{\alpha }\left( \lambda \tilde{A}^{1}\right) \left(
\lambda A^{2}\right) W_{3}^{\alpha }\right\rangle ,
\]%
where the symbol $\sim $ above of $A^{1}$ is to emphasize that the momenta
are equal but the polarizations are different. Now, using the equations of
motion (\ref{Eq of motion})

\[
D_{\alpha }\left( \lambda A\right) =-\left( Q\right) A_{\alpha }+\left(
\lambda \gamma ^{m}\right) _{\alpha }A_{m} ,
\]%
we obtain

\begin{eqnarray*}
\mathcal{A}_{1} &=&+g_{o}^{\prime 2}g_{c}^{\prime }\pi i\alpha ^{\prime
}\left\langle D_{\alpha }\left( \lambda A^{1}\right) \left( \lambda \tilde{A}%
^{1}\right) \left( \lambda A^{2}\right) W^{\alpha }\right\rangle
+g_{o}^{\prime 2}g_{c}^{\prime }\pi i\alpha ^{\prime }\left\langle \left(
\lambda A^{1}\right) D_{\alpha }\left( \lambda \tilde{A}^{1}\right) \left(
\lambda A^{2}\right) W_{3}^{\alpha }\right\rangle  \\
&=&+g_{o}^{\prime 2}g_{c}^{\prime }\pi i\alpha ^{\prime }\left\langle \left[
-\left( Q\right) A_{\alpha }^{1}+\left( \lambda \gamma ^{m}\right) _{\alpha
}A_{m}^{1}\right] \left( \lambda \tilde{A}^{1}\right) \left( \lambda
A^{2}\right) W_{3}^{\alpha }\right\rangle  \\
&&+g_{o}^{\prime 2}g_{c}^{\prime }\pi i\alpha ^{\prime }\left\langle \left(
\lambda A^{1}\right) \left[ -\left( Q\right) \tilde{A}_{\alpha }^{1}+\left(
\lambda \gamma ^{m}\right) _{\alpha }\tilde{A}_{m}^{1}\right] \left( \lambda
A^{2}\right) W_{3}^{\alpha }\right\rangle .
\end{eqnarray*}

Using the fact that an exact BRST term decouples, we obtain

\begin{eqnarray*}
\mathcal{A}_{1} &=&+g_{o}^{\prime 2}g_{c}^{\prime }\pi i\alpha ^{\prime
}\left\langle A_{m}^{1}\left( \lambda \tilde{A}^{1}\right) \left( \lambda
A^{2}\right) \left( \lambda \gamma ^{m}W^{3}\right) \right\rangle  \\
&&+g_{o}^{\prime 2}g_{c}^{\prime }\pi i\alpha ^{\prime }\left\langle \tilde{A%
}_{m}^{1}\left( \lambda A^{1}\right) \left( \lambda A^{2}\right) \left(
\lambda \gamma ^{m}W^{3}\right) \right\rangle  \\
&&-g_{o}^{\prime 2}g_{c}^{\prime }\pi i\alpha ^{\prime }\left\langle
A_{\alpha }^{1}\left( \lambda \tilde{A}^{1}\right) \left( \lambda
A^{2}\right) QW_{3}^{\alpha }\right\rangle  \\
&&+g_{o}^{\prime 2}g_{c}^{\prime }\pi i\alpha ^{\prime }\left\langle \left(
\lambda A^{1}\right) \tilde{A}_{\alpha }^{1}\left( \lambda A^{2}\right)
QW_{3}^{\alpha }\right\rangle ,
\end{eqnarray*}%
and using

\[
QW^{\alpha }=\frac{1}{4}\left( \lambda \gamma ^{mn}\right) ^{\alpha }%
\mathcal{F}_{mn}
\]%
we arrive in

\begin{eqnarray*}
\mathcal{A}_{1} &=&+g_{o}^{\prime 2}g_{c}^{\prime }\pi i\alpha ^{\prime
}\left\langle A_{m}^{1}\left( \lambda \tilde{A}^{1}\right) \left( \lambda
A^{2}\right) \left( \lambda \gamma ^{m}W^{3}\right) \right\rangle  \\
&&+g_{o}^{\prime 2}g_{c}^{\prime }\pi i\alpha ^{\prime }\left\langle \tilde{A%
}_{m}^{1}\left( \lambda A^{1}\right) \left( \lambda A^{2}\right) \left(
\lambda \gamma ^{m}W^{3}\right) \right\rangle  \\
&&-g_{o}^{\prime 2}g_{c}^{\prime }\frac{\pi i\alpha ^{\prime }}{4}%
\left\langle \left( \lambda \gamma ^{mn}A^{1}\right) \left( \lambda \tilde{A}%
^{1}\right) \left( \lambda A^{2}\right) \mathcal{F}_{mn}^{3}\right\rangle  \\
&&+g_{o}^{\prime 2}g_{c}^{\prime }\frac{\pi i\alpha ^{\prime }}{4}%
\left\langle \left( \lambda A^{1}\right) \left( \lambda \gamma ^{mn}\tilde{A}%
^{1}\right) \left( \lambda A^{2}\right) \mathcal{F}_{mn}^{3}\right\rangle .
\end{eqnarray*}

For the last term of the expression (\ref{amplitude ps geral}), we have

\begin{eqnarray}
\mathcal{A}_{2} &=&g_{o}^{\prime 2}g_{c}^{\prime }\int dy_{3}\left\langle
\left( \lambda A^{1}\right) \left( \bar{\lambda}\bar{A}^{1}\right) \left(
\lambda A^{2}\right) \frac{1}{2}N^{mn}F_{mn}^{3}\right\rangle
\label{A1 amplitude ps} \\
&=&g_{o}^{\prime 2}g_{c}^{\prime }\int dy_{3}\frac{\alpha ^{\prime }}{%
8\left( z-y_{3}\right) }\left\langle \left( \lambda \gamma ^{mn}A^{1}\right)
\left( \bar{\lambda}\bar{A}^{1}\right) \left( \lambda A^{2}\right)
F_{mn}^{3}\right\rangle   \nonumber \\
&&+g_{o}^{\prime 2}g_{c}^{\prime }\int dy_{3}\frac{\alpha ^{\prime }}{%
8\left( \bar{z}-y_{3}\right) }\left\langle \left( \lambda A^{1}\right)
\left( \bar{\lambda}\gamma ^{mn}\bar{A}^{1}\right) \left( \lambda
A^{2}\right) F_{mn}^{3}\right\rangle   \nonumber \\
&&+g_{o}^{\prime 2}g_{c}^{\prime }\int dy_{3}\frac{\alpha ^{\prime }}{%
8\left( y_{2}-y_{3}\right) }\left\langle \left( \lambda A^{1}\right) \left(
\bar{\lambda}\bar{A}^{1}\right) \left( \lambda \gamma ^{mn}A^{2}\right)
F_{mn}^{3}\right\rangle ,  \nonumber
\end{eqnarray}%
where we have used the OPE

\[
N^{mn}\left( y_{3}\right) \lambda ^{\alpha }\left( z\right) =\frac{\alpha
^{\prime }}{4\left( z-y_{3}\right) }\left( \lambda \gamma ^{mn}\right)
^{\alpha }.
\]

Fixing above $y_{2}=\infty $ and $z=ia$, we get

\begin{eqnarray}
\mathcal{A}_{2} &=&g_{o}^{\prime 2}g_{c}^{\prime }\frac{\pi i\alpha ^{\prime
}}{4}\left\langle \left( \lambda \gamma ^{mn}A^{1}\right) \left( \bar{\lambda%
}\bar{A}^{1}\right) \left( \lambda A^{2}\right) F_{mn}^{3}\right\rangle
\label{A2 amplitude ps} \\
&&-g_{o}^{\prime 2}g_{c}^{\prime }\frac{\pi i\alpha ^{\prime }}{4}%
\left\langle \left( \lambda A^{1}\right) \left( \bar{\lambda}\gamma ^{mn}%
\bar{A}^{1}\right) \left( \lambda A^{2}\right) F_{mn}^{3}\right\rangle .
\nonumber
\end{eqnarray}

Adding the results (\ref{A1 amplitude ps}) and (\ref{A2 amplitude ps}), we
finally obtain

\begin{eqnarray}
\mathcal{A} &=&g_{o}^{\prime 2}g_{c}^{\prime }\pi i\alpha ^{\prime
}\left\langle A_{m}\left( \lambda \tilde{A}^{1}\right) \left( \lambda
A^{2}\right) \left( \lambda \gamma ^{m}W^{3}\right) \right\rangle
+g_{o}^{\prime 2}g_{c}^{\prime }\pi i\alpha ^{\prime }\left\langle \tilde{A}%
_{m}\left( \lambda A^{1}\right) \left( \lambda A^{2}\right) \left( \lambda
\gamma ^{m}W^{3}\right) \right\rangle   \nonumber \\
&=&g_{o}^{\prime 2}g_{c}^{\prime }\pi i\alpha ^{\prime }\left\langle \left[
A_{m}^{1}\left( \lambda \tilde{A}^{1}\right) +\tilde{A}_{m}^{1}\left(
\lambda A^{1}\right) \right] \left( \lambda A^{2}\right) \left( \lambda
\gamma ^{m}W^{3}\right) \right\rangle .  \label{Amplitude final
ps}
\end{eqnarray}

Although the starting expression (\ref{amplitude inicial ps}) has gauge
invariance, we left the proof to appendix B. At this point, we must expand
the superfields in components and use the measure

\begin{equation}
\langle \left( \lambda \gamma ^{a}\theta \right) \left( \lambda \gamma^{b}\theta \right)
 \left( \lambda \gamma ^{c}\theta \right) \left( \theta
\gamma _{abc}\theta \right) =1  \label{medida}
\end{equation}
in order to find the contribution of each component. The superfield
expansion is given by

\begin{eqnarray*}
\lambda A &=&\frac{1}{2}a_{f}\left( \lambda \gamma ^{f}\theta \right) -\frac{%
1}{3}\left( \xi \gamma _{m}\theta \right) \left( \lambda \gamma ^{m}\theta
\right) -\frac{1}{32}F_{mn}\left( \lambda \gamma _{p}\theta \right) \left(
\theta \gamma ^{mnp}\theta \right) \\
&&+\frac{1}{60}\left( \lambda \gamma _{m}\theta \right) _{\alpha }\left(
\theta \gamma ^{mnp}\theta \right) \left( \partial _{n}\xi \gamma _{p}\theta
\right) ...
\end{eqnarray*}

\begin{equation}
A_{m}=a_{m}-\left( \xi \gamma _{m}\theta \right) -\frac{1}{8}\left( \theta
\gamma _{m}\gamma ^{pq}\theta \right) F_{pq}+\frac{1}{12}\left( \theta
\gamma _{m}\gamma ^{pq}\theta \right) \left( \partial _{p}\xi \gamma
_{q}\theta \right) ...  \label{expansao supercampos}
\end{equation}

\[
\lambda \gamma ^{s}W=\lambda \gamma ^{s}\xi -\frac{1}{4}\left( \lambda
\gamma ^{s}\gamma ^{mn}\theta \right) F_{mn}+\frac{1}{4}\left( \lambda
\gamma ^{s}\gamma ^{mn}\theta \right) \partial _{m}\xi \gamma _{n}\theta +%
\frac{1}{48}\left( \lambda \gamma ^{s}\gamma ^{mn}\theta \right) \left(
\theta \gamma _{n}\gamma ^{pq}\theta \right) \partial _{m}F_{pq}...
\]

These expressions will be used in computations in the next section.

\section{Correlation functions in components}

\subsection{One graviton/dilaton and two photons}

 As explained in the previous section, the vertex operator for
closed strings can be written as the product of the vertex operators of open
strings. First of all, we need to identify the NS-NS contribution in this
product. From our final expression (\ref{Amplitude final ps}), the closed
string contribution is given by

\[
\left[ A_{m}^{1}\left( \lambda \tilde{A}^{1}\right) +\tilde{A}_{m}^{1}\left(
\lambda A^{1}\right) \right] .
\]

Using the superfield expansion, we have the following result for the NS-NS
contribution

\begin{eqnarray}
&&\left( h_{g_1}-\frac{1}{4}\partial _{m_1}h_{h_1}\eta _{g_1t_2}\left( \theta
\gamma ^{t_2}\gamma ^{m_1h_1}\theta \right) \right) \times  \nonumber \\
&&\left( \frac{1}{2}\tilde{h}_{g_2}\left( \lambda \gamma ^{g_2}\theta \right) -%
\frac{1}{16}\partial _{m_1}\tilde{h}_{g_2}\eta _{t_1t_2}\left( \lambda \gamma
^{t_1}\theta \right) \left( \theta \gamma ^{m_1g_2t_2}\theta \right) \right)
\nonumber \\
&&+\left( \tilde{h}_{g_1}-\frac{1}{4}\partial _{m_1}\tilde{h}_{h_1}\eta
_{g_1t_2}\left( \theta \gamma ^{t_2}\gamma ^{m_1h_1}\theta \right) \right) \times
\nonumber \\
&&\left( \frac{1}{2}h_{g_2}\left( \lambda \gamma ^{g_2}\theta \right) -\frac{1%
}{16}\partial _{m_1}h_{g_2}\eta _{t_1t_2}\left( \lambda \gamma ^{t_1}\theta
\right) \left( \theta \gamma ^{m_1g_2t_2}\theta \right) \right)  \nonumber \\
&=&[\frac{1}{2}\left( h_{g_1}\tilde{h}_{g_2}+\tilde{h}_{g_1}h_{g_2}\right)
\left( \lambda \gamma ^{g_2}\theta \right) -\frac{1}{16}\left( h_{g_1}\partial
_{m_1}\tilde{h}_{g_2}+\tilde{h}_{g_1}\partial _{m_1}h_{g_2}\right) \eta
_{t_1t_2}\left( \lambda \gamma ^{t_1}\theta \right) \left( \theta \gamma
^{m_1g_2t_2}\theta \right)  \nonumber \\
&&-\frac{1}{8}\left( \tilde{h}_{g_2}\partial _{m_1}h_{h_1}+h_{g_2}\partial _{m_1}%
\tilde{h}_{h_1}\right) \eta _{g_1t_2}\left( \theta \gamma ^{t_2}\gamma
^{m_1h_1}\theta \right) \left( \lambda \gamma ^{g_2}\theta \right) ] .
\label{nsns identification}
\end{eqnarray}

We must be careful here to identify the NS-NS field, for when we write the
closed string as the product of two open strings, each part must carry half
of the momentum. For example

\begin{eqnarray*}
\left( h_{g_1}\partial _{m_1}\tilde{h}_{g_2}+\tilde{h}_{g_1}\partial
_{m_1}h_{g_2}\right) &=&\left( i\frac{k_{m_1}^{1}}{2}h_{g_1}\tilde{h}_{g_2}+i%
\frac{k_{m_1}^{1}}{2}\tilde{h}_{g_1}h_{g_2}\right) \\
&=&\frac{1}{2}\partial _{m_1}\left( h_{g_1}\tilde{h}_{g_2}+\tilde{h}%
_{g_1}h_{g_2}\right).
\end{eqnarray*}

From this, we can see that, like in Ramond-Neveu-Schwarz, only the symmetric
part of the NS-NS sector contributes. Its traceless part is identified with
the graviton. We will see later how the two form will come from the RR
sector. Now, making the identification

\[
h_{g_1}\tilde{h}_{g_2}+\tilde{h}_{g_1}h_{g_2}=2h_{g_1g_2}
\]
we obtain the graviton contribution

\begin{eqnarray}
&&[h_{g_1g_2}\left( \lambda \gamma ^{g_2}\theta \right) -\frac{1}{16}\partial
_{m_1}h_{g_2g_1}\eta _{t_1t_2}\left( \lambda \gamma ^{t_1}\theta \right) \left(
\theta \gamma ^{m_1g_2t_2}\theta \right)  \label{graviton operator} \\
&&-\frac{1}{8}\partial _{m_1}h_{h_1g_2}\eta _{g_1t_2}\left( \theta \gamma
^{t_2}\gamma ^{m_1h_1}\theta \right) \left( \lambda \gamma ^{g_2}\theta \right)
].  \nonumber
\end{eqnarray}

After the identification of the NS-NS contribution, we must go back to the
expression (\ref{Amplitude final ps}) and consider only the photon
contribution from (\ref{expansao supercampos}) to obtain

\begin{eqnarray*}
\mathcal{A} &=&g_{o}^{\prime 2}g_{c}^{\prime }\pi i\alpha ^{\prime }\langle
\lbrack h_{g_1g_2}\left( \lambda \gamma ^{g_2}\theta \right) -\frac{1}{16}%
\partial _{m_1}h_{g_2g_1}\eta _{t_1t_2}\left( \lambda \gamma ^{t_1}\theta \right)
\left( \theta \gamma ^{m_1g_2t_2}\theta \right) \\
&&-\frac{1}{8}\partial _{m_1}h_{h_1g_2}\eta _{g_1t_2}\left( \theta \gamma
^{t_2}\gamma ^{m_1h_1}\theta \right) \left( \lambda \gamma ^{g_2}\theta \right) ]
\\
&&\left( \frac{1}{2}a_{f_2}^{2}\left( \lambda \gamma ^{f_2}\theta \right) -%
\frac{1}{32}F_{m_2f_2}^{2}\left( \lambda \gamma _{p}\theta \right) \left(
\theta \gamma ^{m_2f_2p}\theta \right) \right) \\
&&\left( -\frac{1}{4}\left( \lambda \gamma ^{g_1}\gamma ^{m_3f_3}\theta \right)
F_{m_3f_3}^{3}+\frac{1}{48}\left( \lambda \gamma ^{g_1}\gamma ^{m_3n}\theta
\right) \left( \theta \gamma _{n}\gamma ^{n_3f_3}\theta \right) \partial
_{m_3}F_{n_3f_3}^{3}\right) \rangle.
\end{eqnarray*}

As we know from eq. (\ref{medida}), only terms with five thetas contribute
to the amplitude. Then, we have

\begin{eqnarray*}
\mathcal{A} &=&g_{o}^{\prime 2}g_{c}^{\prime }\frac{\pi i\alpha ^{\prime }}{%
96}\langle h_{g_1g_2}a_{f_2}^{2}\partial _{m_3}F_{n_3f_3}^{3}\left( \lambda \gamma
^{g_1}\gamma ^{m_3n}\theta \right) \left( \lambda \gamma ^{g_2}\theta \right)
\left( \lambda \gamma ^{f_2}\theta \right) \left( \theta \gamma _{n}\gamma
^{n_3f_3}\theta \right) \rangle \\
&&+g_{o}^{\prime 2}g_{c}^{\prime }\frac{\pi i\alpha ^{\prime }}{128}\langle
h_{g_1g_2}F_{m_2f_2}^{2}F_{m_3f_3}^{3}\left( \lambda \gamma ^{g_1}\gamma
^{m_3f_3}\theta \right) \left( \lambda \gamma ^{g_2}\theta \right) \left(
\lambda \gamma _{p}\theta \right) \left( \theta \gamma ^{m_2f_2p}\theta
\right) \rangle \\
&&+g_{o}^{\prime 2}g_{c}^{\prime }\frac{\pi i\alpha ^{\prime }}{128}\langle
a_{f_2}^{2}\partial _{m_1}\tilde{h}_{g_2g_1}F_{m_3f_3}^{3}\eta _{t_1t_2}\left(
\lambda \gamma ^{g_1}\gamma ^{m_3f_3}\theta \right) \left( \lambda \gamma
^{t_1}\theta \right) \left( \lambda \gamma ^{f_2}\theta \right) \left( \theta
\gamma ^{m_1g_2t_2}\theta \right) \rangle \\
&&+g_{o}^{\prime 2}g_{c}^{\prime }\frac{\pi i\alpha ^{\prime }}{64}\langle
a_{f_2}^{2}\partial _{m_1}h_{g_1g_2}\eta _{t_1t_2}F_{m_3f_3}^{3}\left( \lambda
\gamma ^{t_1}\gamma ^{m_3f_3}\theta \right) \left( \lambda \gamma ^{g_2}\theta
\right) \left( \lambda \gamma ^{f_2}\theta \right) \left( \theta \gamma
^{t_2}\gamma ^{m_1g_1}\theta \right) \rangle.
\end{eqnarray*}

In order to solve the above expression, we need to use the identity (\ref{A4}%
) and successive times the identities (\ref{A1}) and (\ref{A2}) described in
appendix A. Solving term by term with the help of the GAMMA package \cite%
{gamma}\, we obtain

\begin{eqnarray*}
\mathcal{A}_{1} &=&\pi i\alpha ^{\prime }(-\frac{1}{17280}%
h^{g_1g_2}a_{2}^{f_2}\partial ^{g_2}F_{f_2g_1}^{3}-\frac{7}{34560}%
h_{g_1g_2}a_{f_2}^{2}\partial ^{g_1}F_{f_2g_2}^{3}) \\
&=&-\frac{\pi i\alpha ^{\prime }}{3840}h^{g_1g_2}a_{2}^{f_2}\partial
^{g_2}F_{f_2g_1}^{3}=\frac{\pi i\alpha ^{\prime }}{3840}h^{g_1g_2}\partial
^{g_2}a_{2}^{f_2}F_{f_2g_1}^{3}.
\end{eqnarray*}

In a similar way, we can obtain the second term

\[
\mathcal{A}_{2}=-\frac{\pi i\alpha ^{\prime }}{2304}%
h_{g_1}^{g_2}F_{f_3g_2}^{2}F_{3}^{f_3g_1}-\frac{\pi i\alpha ^{\prime }}{23040}%
h_{g_1}^{g_1}F_{f_3g_1}^{2}F_{3}^{f_3g_1}.
\]

The third one is given by

\begin{eqnarray*}
\mathcal{A}_{3} &=&-\frac{\pi i\alpha ^{\prime }}{7680}a_{f_2}^{2}\partial
^{m_1}h^{f_2g_1}F_{g_1m_1}^{3}+\frac{\pi i\alpha ^{\prime }}{23040}\partial
^{m_1}h_{g_1}^{g_1}a_{2}^{g_1}F_{g_1m_1}^{3} \\
&=&\frac{\pi i\alpha ^{\prime }}{7680}\partial
^{m_1}a_{f_2}^{2}h^{f_2g_1}F_{g_1m_1}^{3}-\frac{\pi i\alpha ^{\prime }}{23040}%
h_{g_1}^{g_1}\partial ^{m_1}a_{2}^{g_1}F_{g_1m_1}^{3} \\
&=&\frac{\pi i\alpha ^{\prime }}{7680}h^{f_2g_1}\partial
^{m_1}a_{f_2}^{2}F_{g_1m_1}^{3}+\frac{\pi i\alpha ^{\prime }}{46080}%
h_{g_1}^{g_1}F_{2}^{g_1m_1}F_{g_1m_1}^{3},
\end{eqnarray*}
and the fourth is

\begin{eqnarray*}
\mathcal{A}_{4} &=&-g_{o}^{\prime 2}g_{c}^{\prime }\frac{1}{7680}%
a_{f_2}^{2}\partial ^{m_1}h^{f_2g_2}F_{g_2m_1}^{3} \\
&=&g_{o}^{\prime 2}g_{c}^{\prime }\frac{1}{7680}h^{f_2g_2}\partial
^{m_1}a_{f_2}^{2}F_{g_2m_1}^{3}.
\end{eqnarray*}

Adding all terms, we obtain

\begin{eqnarray*}
\mathcal{A} &=&g_{o}^{\prime 2}g_{c}^{\prime }\pi i\alpha ^{\prime }(\frac{1%
}{3840}h^{g_1g_2}\partial ^{g_2}a_{2}^{f_2}F_{f_2g_1}^{3}-\frac{1}{2304}%
h_{g_1}^{g_2}F_{f_3g_2}^{2}F_{3}^{f_3g_1} \\
&&+\frac{1}{7680}h^{f_2g_1}\partial ^{m_1}a_{f_2}^{2}F_{g_1m_1}^{3}+\frac{1}{7680}%
h^{f_2g_2}\partial ^{m_1}a_{f_2}^{2}F_{g_2m_1}^{3})+ \\
&&-\frac{\pi i\alpha ^{\prime }}{11520}h_{g_1}^{g_1}F_{f_2m_1}^{2}F_{3}^{f_2m_1} \\
&=&g_{o}^{\prime 2}g_{c}^{\prime }\pi i\alpha ^{\prime }(\frac{1}{3840}%
h^{g_1g_2}F_{2}^{g_2f_2}F_{f_2g_1}^{3}-\frac{1}{2304}%
h_{g_1}^{g_2}F_{f_3g_2}^{2}F_{3}^{f_3g_1}) \\
&=&-g_{o}^{\prime 2}g_{c}^{\prime }\frac{\pi i\alpha ^{\prime }}{1440}%
h_{g_2}^{g_1}F_{2}^{g_2f_2}F_{f_2g_1}^{3}-\frac{\pi i\alpha ^{\prime }}{11520}%
h_{g_1}^{g_1}F_{f_2m_1}^{2}F_{3}^{f_2m_1}.
\end{eqnarray*}

In the gauge $k^{m}h_{mn}=0$, the relation $h_{g_1}^{g_1}=4\Phi $ is valid
\cite{polchinski}, and we finally get

\begin{equation}
\mathcal{A}=\frac{\pi i\alpha ^{\prime }}{720}g_{o}^{\prime 2}g_{c}^{\prime
}\left( -\frac{\pi i\alpha ^{\prime }}{2}h_{g_2}^{g_1}F_{2}^{g_2f_2}F_{f_2g_1}^{3}-%
\frac{\pi i\alpha ^{\prime }}{4}\Phi F_{f_2m_1}^{2}F_{3}^{f_2m_1}\right) .
\label{resultado nsns foton}
\end{equation}

Comparing this last expression with (\ref{Acao efetiva supergravidade}), we
see that up to a overall factor we have the right result.

\subsection{One graviton/dilaton and two photinos}

 From the pure spinor viewpoint, we need the graviton and the
photino contribution to the vertex operators which are respectively given by
(\ref{graviton operator}) and (\ref{expansao supercampos}). We then obtain

\begin{eqnarray}
\mathcal{A} &=&g_{o}^{\prime 2}g_{c}^{\prime }\pi i\alpha ^{\prime }\langle
\lbrack h_{g_1g_2}\left( \lambda \gamma ^{g_2}\theta \right) -\frac{1}{16}%
\partial _{m_1}h_{g_2g_1}\eta _{t_1t_2}\left( \lambda \gamma ^{t_1}\theta \right)
\left( \theta \gamma ^{m_1g_2t_2}\theta \right)  \nonumber \\
&&-\frac{1}{8}\partial _{m_1}h_{h_1g_2}\eta _{g_1t_2}\left( \theta \gamma
^{t_2}\gamma ^{m_1h_1}\theta \right) \left( \lambda \gamma ^{g_2}\theta \right) ]
\nonumber \\
&&\left( -\frac{1}{3}\left( \xi ^{2}\gamma _{r}\theta \right) \left( \lambda
\gamma ^{r}\theta \right) +\frac{1}{60}\left( \lambda \gamma _{r}\theta
\right) \left( \theta \gamma ^{rst}\theta \right) \left( \partial _{s}\xi
^{2}\gamma _{t}\theta \right) \right)  \nonumber \\
&&\left( \lambda \gamma ^{m}\xi ^{3}+\frac{1}{4}\left( \lambda \gamma
^{m}\gamma ^{pq}\theta \right) \partial _{p}\xi ^{3}\gamma _{q}\theta
\right) \rangle.  \label{graviton fotino fotino}
\end{eqnarray}

In the expression above, we have four terms with five thetas given by

\begin{eqnarray}
\mathcal{A}=2g_{o}^{\prime 2}g_{c}^{\prime }\pi i\alpha ^{\prime }\langle &&%
\frac{1}{24}h_{mg2}\left( \lambda \gamma ^{m}\gamma ^{m_3f_3}\theta \right)
\left( \lambda \gamma ^{g_2}\theta \right) \left( \lambda \gamma ^{f_2}\theta
\right) \left( \theta \gamma _{f_2}\xi ^{2}\right) \partial _{m_3}\xi
^{3}\gamma _{f_3}\theta \rangle  \label{graviton 2 fotinos exp} \\
&&+2\pi i\alpha ^{\prime }\langle \frac{1}{120}h_{mg2}\left( \lambda \gamma
^{m}\xi ^{3}\right) \left( \partial _{m_2}\xi ^{2}\gamma _{f_2}\theta \right)
\left( \lambda \gamma ^{g_2}\theta \right) \left( \lambda \gamma _{r}\theta
\right) \left( \theta \gamma ^{rm_2f_2}\theta \right) \rangle  \nonumber \\
&&+2\pi i\alpha ^{\prime }\langle \frac{1}{96}\partial _{m_1}h_{g_1g_2}\left(
\lambda \gamma ^{g_1}\xi ^{3}\right) \left( \xi ^{2}\gamma _{f_2}\theta
\right) \left( \lambda \gamma _{p}\theta \right) \left( \lambda \gamma
^{f_2}\theta \right) \left( \theta \gamma ^{m_1g_2p}\theta \right) \rangle
\nonumber \\
&&2\pi i\alpha ^{\prime }\langle \frac{1}{48}\partial
_{m_1}h_{g_1}{}_{g_2}\left( \lambda \gamma _{m}\xi ^{3}\right) \left( \xi
^{2}\gamma _{f_2}\theta \right) \left( \lambda \gamma ^{g_2}\theta \right)
\left( \lambda \gamma ^{f_2}\theta \right) \left( \theta \gamma
^{mm_1g_1}\theta \right) \rangle.  \nonumber
\end{eqnarray}

Using now the identity

\begin{equation}
\xi _{2}^{\alpha }\xi _{3}^{\beta }=\frac{1}{16}\left( \gamma _{a}\right)
^{\alpha \beta }\left( f^{a}\right) +\frac{1}{96}\left( \gamma _{abc}\right)
^{\alpha \beta }\left( f^{abc}\right) +\frac{1}{3840}\left( \gamma
_{abcde}\right) ^{\alpha \beta }\left( f^{abcde}\right),
\label{expansao 2 espinores}
\end{equation}
where

\[
f^{a...}=\xi ^{1}\gamma ^{a...}\xi ^{2},
\]
we have

\[
\xi _{2}^{\alpha }\partial _{m_3}\xi _{3}^{\beta }=\frac{1}{16}\left( \gamma
_{a}\right) ^{\alpha \beta }\left( f_{1m_3}^{a}\right) +\frac{1}{96}\left(
\gamma _{abc}\right) ^{\alpha \beta }\left( f_{1m_3}^{abc}\right) +\frac{1}{%
3840}\left( \gamma _{abcde}\right) ^{\alpha \beta }\left(
f_{1m_3}^{abcde}\right).
\]

Therefore the first term gives

\begin{eqnarray*}
\mathcal{A}_{1} &=&g_{o}^{\prime 2}g_{c}^{\prime }2\pi i\alpha ^{\prime
}\langle \frac{1}{24}h_{g_1g_2}\left( \lambda \gamma ^{g_1}\gamma ^{m_3f_3}\theta
\right) \left( \lambda \gamma ^{g_2}\theta \right) \left( \lambda \gamma
^{f_2}\theta \right) \\
&&\left( \frac{1}{16}\left( \theta \gamma _{f_2}\gamma _{a}\gamma _{f_3}\theta
\right) \left( f_{1m_3}^{a}\right) +\frac{1}{96}\left( \theta \gamma
_{f_2}\gamma _{abc}\gamma _{f_3}\theta \right) \left( f_{1m_3}^{abc}\right) +%
\frac{1}{3840}\left( \theta \gamma _{f_2}\gamma _{abcde}\gamma _{f_3}\theta
\right) \left( f_{1m_3}^{abcde}\right) \right) \rangle.
\end{eqnarray*}

Using now the identities (\ref{A5}),(\ref{A6}) ,(\ref{A7}) and successive
times the identities (\ref{A1}), (\ref{A2}) and (\ref{A3}), we obtain that
only the first term contributes

\begin{eqnarray*}
\mathcal{A}_{1} &=&g_{o}^{\prime 2}g_{c}^{\prime }2\pi i\alpha ^{\prime }%
\left[ -\frac{1}{17280}h_{g_1g_2}\xi ^{2}\gamma ^{g_1}\partial ^{g_2}\xi ^{3}-%
\frac{1}{4320}h_{g_1g_2}\xi _{2}\gamma ^{g_2}\partial ^{g_1}\xi _{3}\right] \\
&=&-g_{o}^{\prime 2}g_{c}^{\prime }\frac{2\pi i\alpha ^{\prime }}{3456}%
h_{g_1g_2}\xi _{2}\gamma ^{g_1}\partial ^{g_2}\xi _{3}.
\end{eqnarray*}

The next term in (\ref{graviton 2 fotinos exp}) is

\[
\mathcal{A}_{2}=g_{o}^{\prime 2}g_{c}^{\prime }\frac{2\pi i\alpha ^{\prime }%
}{120}\langle h_{g_1g_2}\left( \lambda \gamma ^{g_1}\xi ^{3}\right) \left(
\partial _{m_2}\xi ^{2}\gamma _{f_2}\theta \right) \left( \lambda \gamma
^{g_2}\theta \right) \left( \lambda \gamma _{r}\theta \right) \left( \theta
\gamma ^{rm_2f_2}\theta \right) \rangle.
\]

Following the same steps described above, we obtain for the second term

\[
\mathcal{A}_{2}=-g_{o}^{\prime 2}g_{c}^{\prime }\frac{2\pi i\alpha ^{\prime }%
}{17280}h_{g_1g_2}\xi _{3}\gamma ^{g_1}\partial ^{g_2}\xi _{2} .
\]

For the third term we get

\[
\mathcal{A}_{3}=g_{o}^{\prime 2}g_{c}^{\prime }\frac{2\pi i\alpha ^{\prime }%
}{96}\langle \partial _{m_1}h_{g_1g_2}\left( \lambda \gamma ^{g_1}\xi
^{3}\right) \left( \xi ^{2}\gamma _{f_2}\theta \right) \left( \lambda \gamma
_{p}\theta \right) \left( \lambda \gamma ^{f_2}\theta \right) \left( \theta
\gamma ^{m_1g_2p}\theta \right) \rangle,
\]
which has null result. In fact, there is no way to contract a kinetic term
for the graviton with two photinos giving a non null result and

\[
\mathcal{A}_{3}=0.
\]

For the last term we have

\[
\mathcal{A}_{4}=g_{o}^{\prime 2}g_{c}^{\prime }2\pi i\alpha ^{\prime
}\langle \frac{1}{48}\partial _{m_1}h_{g_1}{}_{g_2}\left( \lambda \gamma
_{m}\xi ^{3}\right) \left( \xi ^{2}\gamma _{f_2}\theta \right) \left( \lambda
\gamma ^{g_2}\theta \right) \left( \lambda \gamma ^{f_2}\theta \right) \left(
\theta \gamma ^{mm_1g_1}\theta \right) \rangle,
\]
which is also null for the same reason as before. Finally, adding all terms
we obtain

\begin{eqnarray*}
\mathcal{A} &=&-g_{o}^{\prime 2}g_{c}^{\prime }\frac{\pi i\alpha ^{\prime }}{%
1440}h_{g_1g_2}\xi _{2}\gamma ^{g_1}\partial ^{g_2}\xi _{3} \\
&=&g_{o}^{\prime 2}g_{c}^{\prime }\frac{\pi i\alpha ^{\prime }}{720}\left( -%
\frac{1}{2}h_{g_1g_2}\xi _{2}\gamma ^{g_1}\partial ^{g_2}\xi _{3}\right).
\end{eqnarray*}

In this case the correlation of one dilaton and of two photinos gives a null result
using the photino's equation of motion. The amplitude above is proportional
to the respective term in the effective action (\ref{Acao efetiva
supergravidade}), with the same overall factor of the eq.(\ref{resultado nsns
foton}).

\subsection{One gravitino/dilatino, one photon and one photino}

In the pure spinor computation, we need the gravitino contribution for the
vertex operator. This is given by

\begin{eqnarray*}
A_{\alpha }^{1}\tilde{A}_{m}^{1} &=&-\frac{1}{2}h_{g_1}\left( \gamma
^{g_1}\theta \right) _{\alpha }\left( \tilde{\xi}\gamma _{m}\theta \right) +%
\frac{1}{24}h_{g_1}\left( \gamma ^{g_1}\theta \right) _{\alpha }\left( \theta
\gamma _{m}\gamma ^{pq}\theta \right) \left( \partial _{p}\tilde{\xi}\gamma
_{q}\theta \right)  \\
&&-\frac{1}{3}\tilde{h}_{m}\left( \xi \gamma _{r}\theta \right) \left(
\gamma ^{r}\theta \right) _{\alpha }+\frac{1}{12}\left( \xi \gamma
_{r}\theta \right) \left( \gamma ^{r}\theta \right) _{\alpha }\left( \theta
\gamma _{m}\gamma ^{pq}\theta \right) \partial _{p}\tilde{h}_{q} \\
&&+\frac{1}{16}\partial _{r}h_{s}\left( \gamma _{t}\theta \right) _{\alpha
}\left( \theta \gamma ^{rst}\theta \right) \left( \tilde{\xi}\gamma
_{m}\theta \right) +\frac{1}{60}\tilde{h}_{m}\left( \gamma _{r}\theta
\right) _{\alpha }\left( \theta \gamma ^{rst}\theta \right) \left( \partial
_{s}\xi \gamma _{t}\theta \right) .
\end{eqnarray*}

Then we have
\begin{eqnarray*}
\tilde{A}_{\alpha }^{1}A_{m}^{1}+A_{\alpha }^{1}\tilde{A}_{m}^{1} &=&-\frac{1%
}{2}\left( \gamma ^{g_1}\theta \right) _{\alpha }\left[ \left( h_{g_1}\tilde{%
\xi}+\tilde{h}_{g_1}\xi \right) \gamma _{m}\theta \right]  \\
&&+\frac{1}{24}\left( \gamma ^{g_1}\theta \right) _{\alpha }\left( \theta
\gamma _{m}\gamma ^{pq}\theta \right) \left[ \left( h_{g_1}\partial _{p}%
\tilde{\xi}+\tilde{h}_{g_1}\partial _{p}\xi \right) \gamma _{q}\theta \right]
\\
&&-\frac{1}{3}\left[ \left( h_{m}\tilde{\xi}+\tilde{h}_{m}\xi \right) \gamma
_{r}\theta \right] \left( \gamma ^{r}\theta \right) _{\alpha } \\
&&+\frac{1}{12}\left[ \left( \partial _{p}\tilde{h}_{q}\xi +\partial
_{p}h_{q}\tilde{\xi}\right) \gamma _{r}\theta \right] \left( \gamma
^{r}\theta \right) _{\alpha }\left( \theta \gamma _{m}\gamma ^{pq}\theta
\right)  \\
&&+\frac{1}{16}\left( \gamma _{t}\theta \right) _{\alpha }\left( \theta
\gamma ^{rst}\theta \right) \left[ \left( \partial _{r}h_{s}\tilde{\xi}%
+\partial _{r}\tilde{h}_{s}\xi \right) \gamma _{m}\theta \right]  \\
&&+\frac{1}{60}\left( \gamma _{r}\theta \right) _{\alpha }\left( \theta
\gamma ^{rst}\theta \right) \left[ \left( \tilde{h}_{m}\partial _{s}\xi
+h_{m}\partial _{s}\tilde{\xi}\right) \gamma _{t}\theta \right] .
\end{eqnarray*}

Using the identification

\[
h_{m}\tilde{\xi}+\tilde{h}_{m}\xi =2\psi _{m},\left( h_{g_1}\partial _{p}%
\tilde{\xi}+\tilde{h}_{g_1}\partial _{p}\xi \right) =\partial _{p}\psi _{g_1}
\]

and being careful with the terms with derivatives, we obtain

\begin{eqnarray}
\lambda \tilde{A}^{1}A_{m}^{1}+\lambda A^{1}\tilde{A}_{m}^{1} &=&-\left(
\lambda \gamma ^{g_1}\theta \right) \left( \psi _{g_1}\gamma _{m}\theta
\right) +\frac{1}{24}\left( \lambda \gamma ^{g_1}\theta \right) \left( \theta
\gamma _{m}\gamma ^{pq}\theta \right) \left( \partial _{p}\psi _{g_1}\gamma
_{q}\theta \right)   \nonumber \\
&&-\frac{2}{3}\left( \psi _{m}\gamma _{r}\theta \right) \left( \lambda
\gamma ^{r}\theta \right) +\frac{1}{12}\left( \partial _{p}\psi _{q}\gamma
_{r}\theta \right) \left( \lambda \gamma ^{r}\theta \right) \left( \theta
\gamma _{m}\gamma ^{pq}\theta \right) \label{gravitino cont}   \\
&&+\frac{1}{16}\left( \lambda \gamma _{t}\theta \right) \left( \theta \gamma
^{rst}\theta \right) \left( \partial _{r}\psi _{s}\gamma _{m}\theta \right) +%
\frac{1}{60}\left( \lambda \gamma _{r}\theta \right) \left( \theta \gamma
^{rst}\theta \right) \left( \partial _{m}\psi _{s}\gamma _{t}\theta \right) .
\nonumber
\end{eqnarray}

Now, we go back to the general expression (\ref{Amplitude final ps}) and
consider the contribution of the photon to one of the open strings and the
photino to the other. We obtain

\begin{eqnarray*}
\mathcal{A} &=&g_{o}^{\prime 2}g_{c}^{\prime }\pi i\alpha ^{\prime
}\left\langle \left( \lambda \tilde{A}A_{m}+\lambda A\tilde{A}_{m}\right)
\left( \lambda A^{2}\right) \left( \lambda \gamma ^{m}W\right) \right\rangle
= \\
&=&g_{o}^{\prime 2}g_{c}^{\prime }\pi i\alpha ^{\prime }\langle \lbrack
-\left( \lambda \gamma ^{g_1}\theta \right) \left( \psi _{g_1}\gamma
_{m}\theta \right) -\frac{2}{3}\left( \psi _{m}\gamma _{r}\theta \right)
\left( \lambda \gamma ^{r}\theta \right) \\
&&+\frac{1}{24}\left( \lambda \gamma ^{g_1}\theta \right) \left( \theta
\gamma _{m}\gamma ^{m_1q}\theta \right) \left( \partial _{m_1}\psi _{g_1}\gamma
_{q}\theta \right) +\frac{1}{12}\left( \partial _{m_1}\psi _{g_1}\gamma
_{r}\theta \right) \left( \lambda \gamma ^{r}\theta \right) \left( \theta
\gamma _{m}\gamma ^{m_1g_1}\theta \right) \\
&&+\frac{1}{16}\left( \lambda \gamma _{t}\theta \right) \left( \theta \gamma
^{m_1g_1t}\theta \right) \left( \partial _{m_1}\psi _{g_1}\gamma _{m}\theta
\right) +\frac{1}{60}\left( \lambda \gamma _{r}\theta \right) \left( \theta
\gamma ^{rg1t}\theta \right) \left( \partial _{m}\psi _{g_1}\gamma _{t}\theta
\right) ] \\
&&\left( \frac{1}{2}a_{f_2}^{2}\left( \lambda \gamma ^{f_2}\theta \right) -%
\frac{1}{3}\left( \xi ^{2}\gamma _{t}\theta \right) \left( \lambda \gamma
^{t}\theta \right) -\frac{1}{32}F_{m_2f_2}^{2}\left( \lambda \gamma _{p}\theta
\right) \left( \theta \gamma ^{m_2f_2p}\theta \right) \right) \\
&&\left( \lambda \gamma _{m}\xi ^{3}-\frac{1}{4}\left( \lambda \gamma
_{m}\gamma ^{m_3f_3}\theta \right) F_{m_3f_3}^{3}+\frac{1}{4}\left( \lambda
\gamma _{m}\gamma ^{m_3s}\theta \right) \partial _{m_3}\xi ^{3}\gamma
_{s}\theta +\frac{1}{48}\left( \lambda \gamma _{m}\gamma ^{rs}\theta \right)
\left( \theta \gamma _{s}\gamma ^{m_3f_3}\theta \right) \partial
_{r}F_{m_3f_3}^{3}\right) \rangle.
\end{eqnarray*}

There will be ten terms with five thetas and, as we have two fermions, we
also need to expand them using the identity (\ref{expansao 2 espinores}) to
obtain a total of thirty terms. The details are described in appendix D, and
the result is given by

\[
\mathcal{A}=-g_{o}^{\prime 2}g_{c}^{\prime }\frac{\pi i\alpha ^{\prime }}{720%
}(F_{g_1f_2}^{2}\xi _{3}\gamma ^{f_2}\psi ^{g_1}+F_{m_3f_3}^{3}\xi _{2}\gamma
^{f_3}\psi ^{m_3}).
\]

Again, we get that this is proportional to the respective term of the
effective action (\ref{Acao efetiva supergravidade}) with the right overall
factor. Note that, as in all other terms, this amplitude is symmetric by the
exchange of the two open strings. The dilatino-photino-photon correlation
can be found by this shortcut. We must take the photon contribution from the
fixed operator and the photino from the integrated one. Using the fact that
the amplitude is symmetric by this exchange, we obviously obtain the right
result. We then have

\[
\mathcal{A}=-g_{o}^{\prime 2}g_{c}^{\prime }\frac{\pi i\alpha ^{\prime }}{720%
}(\frac{1}{4}F_{g_1f_2}^{3}\xi _{2}\gamma ^{g_1f_2}\lambda ),
\]
which agree with the desired result.

\subsection{Kalb-Ramond and two photons}

 In the type I superstring, the two form does not come from the
NS-NS sector, as shown before. In fact the two form comes from the RR sector
and only appears as a field strength. The RR contribution to the closed
string vertex operator comes from

\begin{eqnarray*}
\lambda A^{1}\tilde{A}_{m}^{1}+\lambda \tilde{A}^{1}A_{m}^{1} &=&\left( -%
\frac{1}{3}\left( \xi \gamma _{n}\theta \right) \left( \lambda \gamma
^{n}\theta \right) +\frac{1}{60}\left( \lambda \gamma _{m}\theta \right)
_{\alpha }\left( \theta \gamma ^{mnp}\theta \right) \left( \partial _{n}\xi
\gamma _{p}\theta \right) \right) \times  \\
&&\left( -\left( \tilde{\xi}\gamma _{m}\theta \right) +\frac{1}{12}\left(
\theta \gamma _{m}\gamma ^{pq}\theta \right) \left( \partial _{p}\tilde{\xi}%
\gamma _{q}\theta \right) \right)  \\
&&+\left( -\frac{1}{3}\left( \tilde{\xi}\gamma _{n}\theta \right) \left(
\lambda \gamma ^{n}\theta \right) +\frac{1}{60}\left( \lambda \gamma
_{m}\theta \right) _{\alpha }\left( \theta \gamma ^{mnp}\theta \right)
\left( \partial _{n}\tilde{\xi}\gamma _{p}\theta \right) \right) \times  \\
&&\left( -\left( \xi \gamma _{m}\theta \right) +\frac{1}{12}\left( \theta
\gamma _{m}\gamma ^{pq}\theta \right) \left( \partial _{p}\xi \gamma
_{q}\theta \right) \right) .
\end{eqnarray*}

Making the identification

\[
\tilde{\xi}^{a}\xi ^{\beta }+\xi ^{\alpha }\tilde{\xi}^{\beta }=2F^{\alpha
\beta },
\]
we have only one contribution given by

\begin{equation}
\lambda A^{1}\tilde{A}_{m}^{1}+\lambda \tilde{A}^{1}A_{m}^{1}=-\frac{2}{3}%
\left( \lambda \gamma ^{n}\theta \right) \left( \theta \gamma _{m}\right)
_{\alpha }F^{\alpha \beta }\left( \gamma _{n}\theta \right) _{\beta }.
\label{KR operador}
\end{equation}

The other terms have five thetas and do not contribute to the amplitude. We
have then

\begin{eqnarray}
&&\mathcal{A}=g_{o}^{\prime 2}g_{c}^{\prime }\pi i\alpha ^{\prime }\langle
\left( -\frac{2}{3}\left( \lambda \gamma ^{n}\theta \right) \left( \theta
\gamma _{m}\right) _{\alpha }F^{\alpha \beta }\left( \gamma _{n}\theta
\right) _{\beta }\right)  \nonumber \\
&&\left( \frac{1}{2}a_{f_2}^{2}\left( \lambda \gamma ^{f_2}\theta \right) -%
\frac{1}{32}F_{m_2f_2}^{2}\left( \lambda \gamma _{p}\theta \right) \left(
\theta \gamma ^{m_2f_2p}\theta \right) \right)  \nonumber \\
&&\left( -\frac{1}{4}\left( \lambda \gamma ^{g_1}\gamma ^{m_3f_3}\theta \right)
F_{m_3f_3}^{3}+\frac{1}{48}\left( \lambda \gamma ^{g_1}\gamma ^{m_3n}\theta
\right) \left( \theta \gamma _{n}\gamma ^{n_3f_3}\theta \right) \partial
_{m_3}F_{n_3f_3}^{3}\right) \rangle .  \label{amp KR 2
fotons}
\end{eqnarray}

We see that there is just one contribution given by
\[
\mathcal{A}=g_{o}^{\prime 2}g_{c}^{\prime }\frac{\pi i\alpha ^{\prime }}{12}%
\langle \left( \lambda \gamma ^{n}\theta \right) \left( \theta \gamma
_{m}\right) _{\alpha }F^{\alpha \beta }\left( \gamma _{n}\theta \right)
_{\beta }a_{f_2}^{2}\left( \lambda \gamma ^{f_2}\theta \right) \left( \lambda
\gamma ^{m}\gamma ^{pq}\theta \right) F_{pq}^{3}\rangle.
\]

Using now the identity (\ref{expansao 2 espinores}), the RR field can be
expanded

\[
F^{\alpha \beta }=\gamma _{a}^{\alpha \beta }F^{a}+\frac{1}{96}\gamma
_{abc}^{\alpha \beta }H^{abc}+\frac{1}{3840}\gamma _{abcde}^{\alpha \beta
}F^{abcde}.
\]

In the type I superstring the term that survives is the three-form, and we obtain

\[
\mathcal{A}=g_{o}^{\prime 2}g_{c}^{\prime }\frac{\pi i\alpha ^{\prime }}{12}%
\langle H^{abc}F_{pq}^{3}a_{f_2}^{2}\left( \lambda \gamma ^{m}\gamma
^{pq}\theta \right) \left( \lambda \gamma ^{n}\theta \right) \left( \lambda
\gamma ^{f_2}\theta \right) \left( \theta \gamma _{m}\gamma _{abc}\gamma
_{n}\theta \right) \rangle.
\]

In order to solve this term we must use the identities (\ref{A4}), (\ref{A6}%
) and successive applications of the identities (\ref{A1}), (\ref{A2}) and (%
\ref{A3}). The result is

\[
\mathcal{A=}g_{o}^{\prime 2}g_{c}^{\prime }\frac{\pi i\alpha ^{\prime }}{720}%
\left( \frac{1}{8}a_{f_2}^{2}F_{m_3f_3}^{3}H^{f_2f_3m_3}\right),
\]
and it is proportional to the expression (\ref{Acao efetiva supergravidade}%
), as desired. This term is very important because it gives origin to a
coupling which will cancel the mixed anomaly of SO(32) type I superstring.

\subsection{Kalb-Ramond and two photinos}

 The RR contribution to the closed string is given by (\ref{KR
operador}), and we have the amplitude

\begin{eqnarray*}
\mathcal{A} &=&g_{o}^{\prime 2}g_{c}^{\prime }\pi i\alpha ^{\prime }\langle
\left( -\frac{2}{3}\left( \lambda \gamma ^{n}\theta \right) \left( \theta
\gamma _{m}\right) _{\alpha }F^{\alpha \beta }\left( \gamma _{n}\theta
\right) _{\beta }\right) \\
&&\left( -\frac{1}{3}\left( \xi ^{2}\gamma _{r}\theta \right) \left( \lambda
\gamma ^{r}\theta \right) +\frac{1}{60}\left( \lambda \gamma _{r}\theta
\right) \left( \theta \gamma ^{rst}\theta \right) \left( \partial _{s}\xi
^{2}\gamma _{t}\theta \right) \right) \\
&&\left( \lambda \gamma ^{m}\xi ^{3}+\frac{1}{4}\left( \lambda \gamma
^{m}\gamma ^{pq}\theta \right) \partial _{p}\xi ^{3}\gamma _{q}\theta
\right) \rangle.
\end{eqnarray*}

The unique term which has five thetas in the last equation is the following
\begin{eqnarray*}
\mathcal{A} &=&g_{o}^{\prime 2}g_{c}^{\prime }\frac{2\pi i\alpha ^{\prime }}{%
9}\langle \left( \lambda \gamma ^{m}\xi ^{3}\right) \left( \xi ^{2}\gamma
_{t}\theta \right) \left( \lambda \gamma ^{t}\theta \right) \left( \lambda
\gamma ^{n}\theta \right) \left( \theta \gamma \right) _{m\alpha }F^{\alpha
\beta }\left( \gamma _{n}\theta \right) _{\beta }\rangle \\
&=&g_{o}^{\prime 2}g_{c}^{\prime }\frac{2\pi i\alpha ^{\prime }}{9}%
H_{abc}\langle \left( \lambda \gamma ^{m}\xi ^{3}\right) \left( \xi
^{2}\gamma _{t}\theta \right) \left( \lambda \gamma ^{t}\theta \right)
\left( \lambda \gamma ^{n}\theta \right) \left( \theta \gamma _{m}\gamma
_{abc}\gamma _{n}\theta \right) \rangle.
\end{eqnarray*}

We use here the identity (\ref{expansao 2 espinores}) two times and the
identities of the appendix A in order to obtain

\[
\mathcal{A}=-g_{o}^{\prime 2}g_{c}^{\prime }\frac{\pi i\alpha ^{\prime }}{%
34560}H_{abc}\xi ^{2}\gamma ^{abc}\xi ^{3}=g_{o}^{\prime 2}g_{c}^{\prime }%
\frac{\pi i\alpha ^{\prime }}{720}\left( -\frac{1}{48}H_{abc}\xi ^{2}\gamma
^{abc}\xi ^{3}\right)
\]
and this is the right coupling, proportional to eq. (\ref{Acao efetiva
supergravidade}).

\section{Conclusions}

 In this work we have computed explicitly all correlation functions
involving one closed and two open strings in the pure spinor formalism.
Comparing with the effective action for the type I supergravity we came to
the conclusion that the pure spinor formalism survives one more consistency
test and most of the couplings of the effective action were derived here.
The mixed string sector of pure spinor has not previously been considered in the
literature and there is a lot of research yet to be done. The problems
considered here are just the beginning. Higher point amplitudes can be
considered and loop corrections to type I supergravity have not been computed
from the pure spinor viewpoint.

As discussed in this paper, the pure spinor formalism gives the right
coupling between the Kalb-Ramond field and other gauge fields, a result of
particular importance in the mixed anomaly cancellation. At tree level,
diagrams in which a two form is exchanged between two gauge fields on one
side and four on the other side have to be considered for this cancellation.
Therefore, a first step may be the computation of the tree-level five point
amplitudes involving a Kalb-Ramond field and four gauge bosons. This last
idea is left here for future investigations.

\bigskip

\textbf{Acknowledgements}:I would first like to thank my advisor Nathan
Berkovits, for useful discussions. I thank Carlos R. Mafra for clarifying
many things about amplitude computations. I thank the GAMMA package (\cite%
{gamma})of Mathematica and I also acknowledge FAPESP grant 04/06639-4 for
financial support.

\section{Appendix A: Some important identities}

 During the computations made in this paper, we use extensively the
following identities \cite{Berkovits:2006bk}:

\begin{equation}
\langle \left( \lambda \gamma ^{a}\theta \right) \left( \lambda \gamma ^{b}\theta
\right) \left( \lambda \gamma ^{c}\theta \right) \left( \theta \gamma
_{def}\theta \right)\rangle =\frac{1}{120}\delta _{def}^{abc} ,  \label{A1}
\end{equation}

\begin{equation}
\langle \left( \lambda \gamma ^{abc}\theta \right) \left( \lambda \gamma _{d}\theta
\right) \left( \lambda \gamma _{e}\theta \right) \left( \theta \gamma
_{fgh}\theta \right) \rangle =\frac{1}{70}\delta _{\lbrack d}^{[a}\eta _{e][f}\delta
_{g}^{b}\delta _{h]}^{c]},  \label{A2}
\end{equation}

\begin{equation}
\langle \left( \lambda \gamma ^{abcde}\theta \right) \left( \lambda \gamma
_{f}\theta \right) \left( \lambda \gamma_{g}\theta \right) \left(
\theta\gamma _{hij}\theta \right) \rangle =-\frac{1}{42}\delta _{fghij}^{abcde}-%
\frac{1}{5040}\varepsilon _{fghij}^{abcde}.  \label{A3}
\end{equation}

Any other term can be reduced to these above using the identities

\begin{equation}
\gamma ^{a}\gamma ^{bc}=\gamma ^{abc}+\eta ^{ab}\gamma ^{c}-\eta ^{ac}\gamma
^{b},  \label{A4}
\end{equation}

\begin{equation}
\gamma ^{a}\gamma ^{b}\gamma ^{c}=\eta ^{bc}\gamma ^{a}-\eta ^{ac}\gamma
^{b}+\eta ^{ab}\gamma ^{c}+\gamma ^{abc},  \label{A5}
\end{equation}

\begin{eqnarray}
\gamma ^{a}\gamma ^{abc}\gamma ^{d} &=&\eta ^{ad}\eta ^{ce}\gamma ^{b}-\eta
^{ac}\eta ^{de}\gamma ^{b}-\eta ^{ad}\eta ^{be}\gamma ^{c}+\eta ^{ab}\eta
^{de}\gamma ^{c}  \nonumber \\
&&+\eta ^{ac}\eta ^{be}\gamma ^{d}-\eta ^{ab}\eta ^{ce}\gamma ^{d}+\eta
^{de}\gamma ^{abc}-\eta ^{ce}\gamma ^{abd}+\eta ^{be}\gamma ^{acd}  \nonumber
\\
&&-\eta ^{ae}\gamma ^{bcd}+\eta ^{ad}\gamma ^{bce}-\eta ^{ac}\gamma
^{bde}+\eta ^{ab}\gamma ^{cde}+\gamma ^{abcde}  \label{A6}
\end{eqnarray}
and
\begin{eqnarray}
\gamma ^{a}\gamma ^{abcde}\gamma ^{f} &=&\eta ^{af}\eta ^{eg}\gamma
^{bcd}-\eta ^{ae}\eta ^{fg}\gamma ^{bcd}-\eta ^{af}\eta ^{dg}\gamma
^{bce}+\eta ^{ad}\eta ^{fg}\gamma ^{bce}  \nonumber \\
&&+\eta ^{ae}\eta ^{dg}\gamma ^{bcf}-\eta ^{ad}\eta ^{eg}\gamma ^{bcf}+\eta
^{af}\eta ^{cg}\gamma ^{bde}-\eta ^{ac}\eta ^{fg}\gamma ^{bde}  \nonumber \\
&&-\eta ^{ae}\eta ^{cg}\gamma ^{bdf}+\eta ^{ac}\eta ^{eg}\gamma ^{bdf}+\eta
^{ad}\eta ^{cg}\gamma ^{bef}-\eta ^{ac}\eta ^{dg}\gamma ^{bef}  \nonumber \\
&&-\eta ^{af}\eta ^{bg}\gamma ^{cde}+\eta ^{ab}\eta ^{fg}\gamma ^{cde}+\eta
^{ae}\eta ^{bg}\gamma ^{cdf}-\eta ^{ab}\eta ^{eg}\gamma ^{cdf}  \nonumber \\
&&-\eta ^{ad}\eta ^{bg}\gamma ^{cef}+\eta ^{ab}\eta ^{dg}\gamma ^{cef}+\eta
^{ac}\eta ^{bg}\gamma ^{def}-\eta ^{ab}\eta ^{cg}\gamma ^{def}  \nonumber \\
&&+\eta ^{fg}\gamma ^{abcde}-\eta ^{eg}\gamma ^{abcdf}+\eta ^{dg}\gamma
^{abcef}-\eta ^{cg}\gamma ^{abdef}  \nonumber \\
&&+\eta ^{bg}\gamma ^{acdef}-\eta ^{ag}\gamma ^{bcdef}+\eta ^{af}\gamma
^{bcdeg}-\eta ^{ae}\gamma ^{bcdfg}  \nonumber \\
&&+\eta ^{ad}\gamma ^{bcefg}-\eta ^{ac}\gamma ^{bdefg}+\eta ^{ab}\gamma
^{cdefg}+\gamma ^{abcdefg},  \label{A7}
\end{eqnarray}

\section{Appendix B: Gauge invariance}

 The expression (\ref{Amplitude final ps}) must be invariant under
all gauge transformations. The first is given by

\[
\delta \left( \lambda A^{2}\right) =Q\Lambda
\]
and the variation of the first term in (\ref{Amplitude final ps}) is

\begin{eqnarray*}
\frac{\delta S_{1}}{\pi i\alpha ^{\prime }} &=&\left\langle A_{m}^{1}\left(
\lambda \tilde{A}^{1}\right) \left( Q\Lambda \right) \left( \lambda \gamma
^{m}W^{3}\right) \right\rangle \\
&=&\left\langle QA_{m}^{1}\left( \lambda \tilde{A}^{1}\right) \Lambda \left(
\lambda \gamma ^{m}W^{3}\right) \right\rangle +\left\langle A_{m}^{1}Q\left(
\lambda \tilde{A}^{1}\right) \Lambda \left( \lambda \gamma ^{m}W^{3}\right)
\right\rangle -\left\langle A_{m}^{1}\left( \lambda \tilde{A}^{1}\right)
\Lambda Q\left( \lambda \gamma ^{m}W^{3}\right) \right\rangle \\
&=&\left\langle \left[ \lambda \gamma _{m}W+\partial _{m}\left( \lambda
A^{1}\right) \right] \left( \lambda \tilde{A}^{1}\right) \Lambda \left(
\lambda \gamma ^{m}W^{3}\right) \right\rangle -\frac{1}{4}\left\langle
A_{m}^{1}\left( \lambda \tilde{A}^{1}\right) \Lambda \left( \left( \lambda
\gamma ^{m}\right) _{\alpha }\left( \lambda \gamma ^{rs}\right) ^{\alpha
}F_{rs}^{3}\right) \right\rangle \\
&=&\left\langle \left[ \lambda \gamma _{m}W+\partial _{m}\left( \lambda
A^{1}\right) \right] \left( \lambda \tilde{A}^{1}\right) \Lambda \left(
\lambda \gamma ^{m}W^{3}\right) \right\rangle -\frac{1}{4}\left\langle
A_{m}^{1}\left( \lambda \tilde{A}^{1}\right) \Lambda \left( \left( \lambda
\gamma ^{m}\right) _{\alpha }\left( \lambda \gamma ^{rs}\right) ^{\alpha
}F_{rs}^{3}\right) \right\rangle \\
&=&\left\langle \partial _{m}\left( \lambda A^{1}\right) \left( \lambda
\tilde{A}^{1}\right) \Lambda \left( \lambda \gamma ^{m}W^{3}\right)
\right\rangle =k_{m}^{1}\left\langle \left( \lambda A^{1}\right) \left(
\lambda \tilde{A}^{1}\right) \Lambda \left( \lambda \gamma ^{m}W^{3}\right)
\right\rangle.
\end{eqnarray*}

In the above expression, we have used the pure spinor condition (\ref%
{vinculo}) and the Fierz identity
\[
\left( \gamma _{m}\right) _{(\alpha \beta }\left( \gamma ^{m}\right) _{\rho
)\sigma }=0.
\]

The variation of the second term is

\begin{eqnarray*}
\frac{\delta S_{2}}{\pi i\alpha ^{\prime }} &=&\left\langle \tilde{A}%
_{m}^{1}\left( \lambda A^{1}\right) \left( Q\Lambda \right) \left( \lambda
\gamma ^{m}W^{3}\right) \right\rangle  \\
&=&\left\langle Q\tilde{A}_{m}^{1}\left( \lambda A^{1}\right) \Lambda \left(
\lambda \gamma ^{m}W^{3}\right) \right\rangle +\left\langle \tilde{A}%
_{m}^{1}Q\left( \lambda A^{1}\right) \Lambda \left( \lambda \gamma
^{m}W^{3}\right) \right\rangle -\left\langle \tilde{A}_{m}^{1}\left( \lambda
A^{1}\right) \Lambda Q\left( \lambda \gamma ^{m}W^{3}\right) \right\rangle
\\
&=&\left\langle \left[ \lambda \gamma _{m}W^{1}+\partial _{m}\left( \lambda
\tilde{A}^{1}\right) \right] \left( \lambda A^{1}\right) \Lambda \left(
\lambda \gamma ^{m}W^{3}\right) \right\rangle -\frac{1}{4}\left\langle
\tilde{A}_{m}^{1}\left( \lambda A^{1}\right) \Lambda \left( \lambda \gamma
^{m}\lambda \gamma ^{rs}F_{rs}^{3}\right) \right\rangle  \\
&=&\left\langle \left[ \lambda \gamma _{m}W^{1}+\partial _{m}\left( \lambda
\tilde{A}^{1}\right) \right] \left( \lambda A^{1}\right) \Lambda \left(
\lambda \gamma ^{m}W^{3}\right) \right\rangle -\frac{1}{4}\left\langle
\tilde{A}_{m}^{1}\left( \lambda A^{1}\right) \Lambda \left( \lambda \gamma
^{m}\lambda \gamma ^{rs}F_{rs}^{3}\right) \right\rangle  \\
&=&\left\langle \partial _{m}\left( \lambda \tilde{A}^{1}\right) \left(
\lambda A^{1}\right) \Lambda \left( \lambda \gamma ^{m}W^{3}\right)
\right\rangle =k_{m}^{1}\left\langle \left( \lambda \tilde{A}^{1}\right)
\left( \lambda A^{1}\right) \Lambda \left( \lambda \gamma ^{m}W^{3}\right)
\right\rangle  \\
&=&-k_{m}^{1}\left\langle \left( \lambda A^{1}\right) \left( \lambda \tilde{A%
}^{1}\right) \Lambda \left( \lambda \gamma ^{m}W^{3}\right) \right\rangle
\end{eqnarray*}%
again we have used the pure spinor condition and the Fierz identity. Adding
the results we obtain

\[
\delta S=\delta S_{1}+\delta S_{2}=0.
\]

The other gauge transformation is given by

\[
\delta \left( \lambda A^{1}\right) =Q\Lambda ,\delta A_{m}^{1}=\partial
_{m}\Lambda
\]
and we obtain

\[
\frac{\delta S_{1}}{\pi i\alpha ^{\prime }}=\left\langle \partial
_{m}\Lambda \left( \lambda \tilde{A}^{1}\right) \left( \lambda A^{2}\right)
\left( \lambda \gamma ^{m}W^{3}\right) \right\rangle =k_{m}^{1}\left\langle
\Lambda \left( \lambda \tilde{A}^{1}\right) \left( \lambda A^{2}\right)
\left( \lambda \gamma ^{m}W^{3}\right) \right\rangle.
\]

For the second term

\begin{eqnarray*}
\frac{\delta S_{2}}{\pi i\alpha ^{\prime }} &=&\left\langle \tilde{A}%
_{m}^{1}\left( Q\Lambda \right) \left( \lambda A^{2}\right) \left( \lambda
\gamma ^{m}W^{3}\right) \right\rangle  \\
&=&-\left\langle Q\tilde{A}_{m}^{1}\Lambda \left( \lambda A^{2}\right)
\left( \lambda \gamma ^{m}W^{3}\right) \right\rangle +\left\langle \tilde{A}%
_{m}^{1}\Lambda Q\left( \lambda A^{2}\right) \left( \lambda \gamma
^{m}W^{3}\right) \right\rangle -\left\langle \tilde{A}_{m}^{1}\Lambda \left(
\lambda A^{2}\right) Q\left( \lambda \gamma ^{m}W^{3}\right) \right\rangle
\\
&=&-\left\langle \left[ \lambda \gamma _{m}W^{1}+\partial _{m}\left( \lambda
\tilde{A}^{1}\right) \right] \Lambda \left( \lambda A^{2}\right) \left(
\lambda \gamma ^{m}W^{3}\right) \right\rangle -\frac{1}{4}\left\langle
\tilde{A}_{m}^{1}\Lambda \left( \lambda A^{2}\right) \left( \lambda \gamma
^{m}\lambda \gamma ^{rs}F_{rs}^{3}\right) \right\rangle  \\
&=&-\left\langle \left[ \lambda \gamma _{m}W^{1}+\partial _{m}\left( \lambda
\tilde{A}^{1}\right) \right] \Lambda \left( \lambda A^{2}\right) \left(
\lambda \gamma ^{m}W^{3}\right) \right\rangle -\frac{1}{4}\left\langle
\tilde{A}_{m}^{1}\Lambda \left( \lambda A^{2}\right) \left( \lambda \gamma
^{m}\lambda \gamma ^{rs}F_{rs}^{3}\right) \right\rangle  \\
&=&-\left\langle \partial _{m}\left( \lambda \tilde{A}^{1}\right) \Lambda
\left( \lambda A^{2}\right) \left( \lambda \gamma ^{m}W^{3}\right)
\right\rangle =-k_{m}^{1}\left\langle \left( \lambda \tilde{A}^{1}\right)
\Lambda \left( \lambda A^{2}\right) \left( \lambda \gamma ^{m}W^{3}\right)
\right\rangle
\end{eqnarray*}%
and finally

\[
\delta S=\delta S_{1}+\delta S_{2}=0.
\]

Using identical arguments as above we obtain the invariance under

\[
\delta \lambda \tilde{A}=Q\Lambda.
\]

Therefore, as expected, the final expression is in fact gauge invariant.

\section{Appendix C: The one graviton two photons correlation function in
Ramond-Neveu-Schwarz formalism}

 In the Ramond-Neveu-Schwarz case two of the vertex operators must
be in the picture $-1$ and one in the picture $0$. Choosing the closed
string in the picture $0$, we obtain

\[
V_{c}^{0}=\frac{-2i}{\alpha ^{\prime }}g_{c}^{\prime }:c\widetilde{c}h_{\mu
\nu }\left( i\partial X^{\mu }+\frac{\alpha \prime }{2}k_{\sigma }^{1}\psi
^{\sigma }\psi ^{\mu }\right) \left( i\bar{\partial}X^{\nu }+\frac{\alpha
\prime }{2}k_{\rho }^{1}\bar{\psi}^{\rho }\bar{\psi}^{\nu }\right)
e^{ik^{1}\cdot x}(z):.
\]

In this section, we follow the notation used in (\cite{polchinski}).The
fixed open string operator in the -1 picture is given by

\[
V_{o}^{-1}=ig_{o}^{\prime }:a_{2\alpha }\psi ^{\alpha }ce^{-\phi
}e^{ik_{2}\cdot x}(y_{2}):,
\]
and the integrated one is given by
\[
V_{o}^{-1}=ig_{o}^{\prime }\int dy_{3}:a_{3\beta }\psi ^{\beta }e^{-\phi
}e^{ik_{3}\cdot x}(y_{3}):.
\]

The expression for the amplitude is given by (\cite{polchinski})

\begin{eqnarray}
\mathcal{A} &=&2i\frac{g_{c}^{\prime }}{\alpha \prime }g_{o}^{\prime
2}e^{-\lambda }\int_{-\infty }^{+\infty }dy_{3}\langle :c\widetilde{c}h_{\mu
\nu }\left( i\partial X^{\mu }+\frac{\alpha \prime }{2}k_{\sigma }^{1}\psi
^{\sigma }\psi ^{\mu }\right) \left( i\bar{\partial}X^{\nu }+\frac{\alpha
\prime }{2}k_{\rho }^{1}\bar{\psi}^{\rho }\bar{\psi}^{\nu }\right)
e^{ik^{1}\cdot x}(z):  \nonumber \\
&:&a_{2\alpha }\psi ^{\alpha }ce^{-\phi }e^{ik_{2}\cdot x}(y_{2})::a_{3\beta
}\psi ^{\beta }e^{-\phi }e^{ik_{3}\cdot x}(y_{3}):\rangle .
\label{Amplitude RNS}
\end{eqnarray}

The OPEs between the $X^{\mu }$ fields will be needed for all cases and it
is given by
\begin{equation}
:X^{\;\mu }\left( z_{1}\right) X^{\nu }\left( z_{2}\right) :=X^{\mu }\left(
z_{1}\right) X^{\nu }\left( z_{2}\right) -\frac{\alpha \prime }{2}\eta ^{\mu
\nu }\left[ \ln \left\vert z_{1}-z_{2}\right\vert ^{2}+\ln \left\vert z_{1}-%
\bar{z}_{2}\right\vert ^{2}\right] .  \label{ope x}
\end{equation}

From the above expression all the related OPEs can be obtained

\[
:\partial X^{\;\mu }\left( z_{1}\right) \bar{\partial}X^{\nu }\left(
z_{2}\right) :=\partial X^{\mu }\left( z_{1}\right) \bar{\partial}X^{\nu
}\left( z_{2}\right) -\frac{\alpha \prime }{2}\eta ^{\mu \nu }\frac{1}{%
\left( \bar{z}_{2}-z_{1}\right) ^{2}},
\]

\[
:\partial X^{\;\mu }\left( z_{1}\right) X^{\nu }\left( y\right) :=\partial
X^{\mu }\left( z_{1}\right) X^{\nu }\left( y\right) +\alpha \prime \eta
^{\mu \nu }\frac{1}{y-z_{1}},
\]

\[
:\bar{\partial}X^{\;\mu }\left( z_{1}\right) X^{\nu }\left( y\right)
:=\partial X^{\mu }\left( z_{1}\right) X^{\nu }\left( z_{2}\right) +\alpha
\prime \eta ^{\mu \nu }\frac{1}{y-\bar{z}_{1}}.
\]

We also need of the OPE for the fields

\[
\langle e^{-\phi }\left( z_{1}\right) e^{-\phi }(z_{2})\rangle =z_{12}^{-1}
,
\]

\[
\langle \psi ^{\mu }\left( z_{1}\right) \psi ^{\nu }(z_{2})\rangle =\eta
^{\mu \nu }z_{12}^{-1}.
\]

After making all the possible contractions in the expression (\ref{Amplitude
RNS}), we obtain

\begin{eqnarray}
\mathcal{A} &=&2i\frac{g\prime _{c}}{\alpha \prime }g_{o}^{\prime
2}e^{-\lambda }\int_{-\infty }^{+\infty }dy_{3}\langle :c\widetilde{c}%
e^{ik^{1}\cdot x}(z)::ce^{-\phi }e^{ik_{2}\cdot x}(y_{2})::e^{ik_{3}\cdot
x}(y_{3}):\rangle \frac{1}{y_{2}-y_{3}}  \nonumber \\
&&\{-h_{\mu \nu }a_{2\alpha }a_{3\beta }\frac{\eta ^{\alpha \beta }}{\left(
y_{2}-y_{3}\right) }\left[ +\frac{i\alpha \prime }{y_{2}-z}k_{2}^{\mu }+%
\frac{ik_{3}^{\mu }\alpha \prime }{y_{3}-z}\right] \left[ \frac{i\alpha
\prime }{y_{2}-\overline{z}}k_{2}^{\nu }+\frac{i\alpha \prime }{y_{3}-%
\overline{z}}k_{3}^{\nu }\right]  \label{amplitude 1 graviton 2 fotons} \\
&&+\frac{i\alpha ^{\prime }}{2}h_{\mu \nu }\left[ \frac{i\alpha \prime }{%
y_{2}-\overline{z}}k_{2}^{\nu }+\frac{i\alpha \prime }{y_{3}-\overline{z}}%
k_{3}^{\nu }\right] \left[ \frac{k_{1\sigma }\eta ^{\mu \alpha }\eta
^{\sigma \beta }a_{2\alpha }a_{3\beta }}{\left( z-y_{2}\right) \left(
z-y_{3}\right) }-\frac{k_{1\sigma }\eta ^{\mu \beta }\eta ^{\sigma \alpha
}a_{2\alpha }a_{3\beta }}{\left( z-y_{2}\right) \left( z-y_{3}\right) }%
\right]  \nonumber \\
&&+\frac{i\alpha ^{\prime }}{2}h_{\mu \nu }\left[ +\frac{i\alpha \prime }{%
y_{2}-z}k_{2}^{\mu }+\frac{ik_{3}^{\mu }\alpha \prime }{y_{3}-z}\right] %
\left[ \frac{k_{1\rho }\eta ^{\nu \alpha }\eta ^{\rho \beta }a_{2\alpha
}a_{3\beta }}{\left( \bar{z}-y_{2}\right) \left( \bar{z}-y_{3}\right) }-%
\frac{k_{1\rho }\eta ^{\nu \beta }\eta ^{\rho \alpha }a_{2\alpha }a_{3\beta }%
}{\left( \bar{z}-y_{2}\right) \left( \bar{z}-y_{3}\right) }\right] \} .
\nonumber
\end{eqnarray}

The ghost contribution to the amplitude is given by

\[
\left\langle c\widetilde{c}(z)c(z_{2})\right\rangle =C_{D_{2}}^{g}\left\vert
y_{2}-z\right\vert ^{2}\left( z-\overline{z}\right) .
\]

In the last equation, $C_{D_{2}}^{g}$ is a constant coming from functional
determinants. The contribution from the exponentials is given by

\begin{eqnarray*}
&&\left\langle :e^{ik\cdot x}(z)::e^{ik\cdot x}(y_{2})::e^{ik\cdot
x}(y_{3}):\right\rangle \\
&=&iC_{D_{2}}^{X}(2\pi )^{d}\delta (\Sigma k)\left\vert z-\overline{z}%
\right\vert ^{\alpha \prime k_{1}^{2}/2}\left\vert y_{2}-y_{3}\right\vert
^{2\alpha \prime k_{2}\cdot k_{3}}\left\vert y_{2}-z\right\vert ^{2\alpha
\prime k_{1}\cdot k_{2}}\left\vert y_{3}-z\right\vert ^{2\alpha \prime
k_{1}\cdot k_{3}}
\end{eqnarray*}
again, $C_{D_{2}}^{X}$ is a constant coming from functional determinants.
Using momentum conservation we obtain

\[
k_{1}^{2}=k_{1}\cdot k_{2}=k_{1}\cdot k_{3}=k_{3}\cdot k_{2}=0 ,
\]
then

\[
\left\langle :e^{ik\cdot x}(z)::e^{ik\cdot x}(y_{2})::e^{ik\cdot
x}(y_{3}):\right\rangle =iC_{D_{2}}^{X}(2\pi )^{d}\delta (\Sigma k) ,
\]
and we obtain for (\ref{amplitude 1 graviton 2 fotons})

\begin{eqnarray*}
\mathcal{A} &=&-2\frac{g_{c}^{\prime }}{\alpha \prime }g_{o}^{\prime 2}(2\pi
)^{d}\delta (\Sigma k)e^{-\lambda }C_{D_{2}}^{g}C_{D_{2}}^{X}\int_{-\infty
}^{+\infty }dy_{3}\frac{1}{y_{2}-y_{3}}\left\vert y_{2}-z\right\vert
^{2}\left( z-\overline{z}\right) (2\pi )^{d}\delta (\Sigma k) \\
&&\{-h_{\mu \nu }a_{2\alpha }a_{3\beta }\frac{\eta ^{\alpha \beta }}{\left(
y_{2}-y_{3}\right) }\left[ +\frac{i\alpha \prime }{y_{2}-z}k_{2}^{\mu }+%
\frac{ik_{3}^{\mu }\alpha \prime }{y_{3}-z}\right] \left[ \frac{i\alpha
\prime }{y_{2}-\overline{z}}k_{2}^{\nu }+\frac{i\alpha \prime }{y_{3}-%
\overline{z}}k_{3}^{\nu }\right] \\
&&+\frac{i\alpha ^{\prime }}{2}h_{\mu \nu }\left[ \frac{i\alpha \prime }{%
y_{2}-\overline{z}}k_{2}^{\nu }+\frac{i\alpha \prime }{y_{3}-\overline{z}}%
k_{3}^{\nu }\right] \left[ \frac{k_{1\sigma }\eta ^{\mu \alpha }\eta
^{\sigma \beta }a_{2\alpha }a_{3\beta }}{\left( z-y_{2}\right) \left(
z-y_{3}\right) }-\frac{k_{1\sigma }\eta ^{\mu \beta }\eta ^{\sigma \alpha
}a_{2\alpha }a_{3\beta }}{\left( z-y_{2}\right) \left( z-y_{3}\right) }%
\right] \\
&&+\frac{i\alpha ^{\prime }}{2}h_{\mu \nu }\left[ +\frac{i\alpha \prime }{%
y_{2}-z}k_{2}^{\mu }+\frac{ik_{3}^{\mu }\alpha \prime }{y_{3}-z}\right] %
\left[ \frac{k_{1\rho }\eta ^{\nu \alpha }\eta ^{\rho \beta }a_{2\alpha
}a_{3\beta }}{\left( \bar{z}-y_{2}\right) \left( \bar{z}-y_{3}\right) }-%
\frac{k_{1\rho }\eta ^{\nu \beta }\eta ^{\rho \alpha }a_{2\alpha }a_{3\beta }%
}{\left( \bar{z}-y_{2}\right) \left( \bar{z}-y_{3}\right) }\right] \}.
\end{eqnarray*}

The contribution of the functional determinants can be found in \cite%
{polchinski} and it is given by

\[
e^{-\lambda }C_{D_{2}}^{g}C_{D_{2}}^{X}=\frac{1}{\alpha ^{\prime }g_{o}^{2}}%
,g_{c}^{\prime }=\frac{2g_{c}}{\alpha \prime };g_{o}^{\prime }=\frac{g_{o}}{%
\sqrt{2\alpha ^{\prime }}} .
\]

As in the pure spinor case, we fix

\[
y_{2}=0;\mbox{Re}(z)=0;\mbox{Im}(z)=a
\]
to obtain

\begin{eqnarray*}
\mathcal{A} &=&\frac{-i}{2\alpha ^{\prime }}g_{c}^{\prime }(2\pi )^{d}\delta
^{d}(\Sigma k)a\int_{-\infty }^{+\infty }dy_{3}\frac{1}{\left\vert
y_{3}+ia\right\vert ^{2}} \\
&&\left\{ \left[ a_{3}\cdot k_{12}h_{\mu \nu }\left( a_{2}^{\mu }k_{23}^{\nu
}+a_{2}^{\nu }k_{23}^{\mu }\right) -a_{2}\cdot k_{13}h_{\mu \nu }\left(
a_{3}^{\mu }k_{23}^{\nu }+a_{3}^{\nu }k_{23}^{\mu }\right) +2h_{\mu \nu
}k_{23}^{\nu }k_{23}^{\mu }a_{2}\cdot a_{3}\right] \right\}.
\end{eqnarray*}

From the above expression, we can already see that the antisymmetric part of
$h_{\mu \nu }$ does not contribute for this amplitude. In fact in the type I
superstring, the Kalb-Ramond contribution comes from the RR sector and not
from NS-NS. Finally, integrating we obtain

\[
\mathcal{A}=\frac{\pi i}{2\alpha \prime }g_{c}(2\pi )^{d}\delta ^{d}(\Sigma
k)\left[ a_{3}\cdot k_{12}h_{\mu \nu }a_{2}^{\mu }k_{23}^{\nu }-a_{2}\cdot
k_{13}h_{\mu \nu }a_{3}^{\mu }k_{23}^{\nu }+h_{\mu \nu }k_{23}^{\nu
}k_{23}^{\mu }a_{2}\cdot a_{3}\right] .
\]

The last expression can be written in the position space

\begin{equation}
\mathcal{A}=\frac{i}{4\alpha \prime }g_{c}h_{\nu }^{\mu }F_{\mu \alpha
}F^{\nu \alpha }.  \label{resultado RNS}
\end{equation}

This amplitude originates a term in the effective action that is
proportional to (\ref{Acao efetiva supergravidade}) and to the pure spino
result (\ref{resultado nsns foton}),as desired. Obviously it has all the
desired properties as gauge invariance and symmetry in the exchange of the
two photons.

\section{Appendix D: One gravitino one photon one photino}

As said in the text, the final expression is given by

\begin{eqnarray*}
\mathcal{A} &=&-\frac{\pi i\alpha ^{\prime }}{8}\langle \lbrack \left(
\lambda \gamma ^{g_1}\theta \right) \left( \psi _{g_1}\gamma _{m}\theta
\right) +\frac{2}{3}\left( \psi _{m}\gamma _{r}\theta \right) \left( \lambda
\gamma ^{r}\theta \right) ]a_{f_2}^{2}\left( \lambda \gamma ^{f_2}\theta
\right) \left( \lambda \gamma _{m}\gamma ^{m_3s}\theta \right) \partial
_{m_3}\xi ^{3}\gamma _{s}\theta \rangle \\
&&-\frac{\pi i\alpha ^{\prime }}{12}\langle \lbrack \left( \lambda \gamma
^{g_1}\theta \right) \left( \psi _{g_1}\gamma _{m}\theta \right) +\frac{2}{3}%
\left( \psi _{m}\gamma _{r}\theta \right) \left( \lambda \gamma ^{r}\theta
\right) ]\left( \xi ^{2}\gamma _{t}\theta \right) \left( \lambda \gamma
^{t}\theta \right) \left( \lambda \gamma _{m}\gamma ^{m_3f_3}\theta \right)
F_{m_3f_3}^{3}\rangle \\
&&+\frac{\pi i\alpha ^{\prime }}{32}\langle \lbrack \left( \lambda \gamma
^{g_1}\theta \right) \left( \psi _{g_1}\gamma _{m}\theta \right) +\frac{2}{3}%
\left( \psi _{m}\gamma _{r}\theta \right) \left( \lambda \gamma ^{r}\theta
\right) ]F_{m_2f_2}^{2}\left( \lambda \gamma _{p}\theta \right) \left( \theta
\gamma ^{m_2f_2p}\theta \right) \left( \lambda \gamma _{m}\xi ^{3}\right)
\rangle \\
&&+\pi i\alpha ^{\prime }\langle \lbrack \frac{1}{48}\left( \lambda \gamma
^{g_1}\theta \right) \left( \theta \gamma _{m}\gamma ^{m_1q}\theta \right)
\left( \partial _{m_1}\psi _{g_1}\gamma _{q}\theta \right) a_{f_2}^{2}\left(
\lambda \gamma ^{f_2}\theta \right) \left( \lambda \gamma _{m}\xi ^{3}\right)
\\
&&+\frac{1}{24}\left( \partial _{m_1}\psi _{g_1}\gamma _{r}\theta \right)
\left( \lambda \gamma ^{r}\theta \right) \left( \theta \gamma _{m}\gamma
^{m_1g_1}\theta \right) a_{f_2}^{2}\left( \lambda \gamma ^{f_2}\theta \right)
\left( \lambda \gamma _{m}\xi ^{3}\right) \\
&&+\frac{1}{32}\left( \lambda \gamma _{t}\theta \right) \left( \theta \gamma
^{m_1g_1t}\theta \right) \left( \partial _{m_1}\psi _{g_1}\gamma _{m}\theta
\right) a_{f_2}^{2}\left( \lambda \gamma ^{f_2}\theta \right) \left( \lambda
\gamma _{m}\xi ^{3}\right) \\
&&+\frac{1}{120}\left( \lambda \gamma _{r}\theta \right) \left( \theta
\gamma ^{rg1t}\theta \right) \left( \partial _{m}\psi _{g_1}\gamma _{t}\theta
\right) a_{f_2}^{2}\left( \lambda \gamma ^{f_2}\theta \right) \left( \lambda
\gamma _{m}\xi ^{3}\right) ]\rangle
\end{eqnarray*}
the terms with five $\theta $'s are given by

\begin{eqnarray*}
\mathcal{A} &=&+\frac{\pi i\alpha ^{\prime }}{8}a_{f_2}^{2}\langle \left(
\lambda \gamma _{r}\gamma ^{m_3s}\theta \right) \left( \lambda \gamma
^{g_1}\theta \right) \left( \lambda \gamma ^{f_2}\theta \right) \left( \theta
\gamma ^{r}\psi _{g_1}\right) \partial _{m_3}\xi ^{3}\gamma _{s}\theta \rangle
\\
&&+\frac{\pi i\alpha ^{\prime }}{12}a_{f_2}^{2}\langle \left( \lambda \gamma
^{g_1}\gamma ^{m_3s}\theta \right) \left( \lambda \gamma ^{r}\theta \right)
\left( \lambda \gamma ^{f_2}\theta \right) \left( \theta \gamma _{r}\psi
_{g_1}\right) \partial _{m_3}\xi ^{3}\gamma _{s}\theta \rangle \\
&&+\frac{\pi i\alpha ^{\prime }}{12}F_{m_3f_3}^{3}\langle \left( \lambda
\gamma ^{r}\gamma ^{m_3f_3}\theta \right) \left( \lambda \gamma ^{g_1}\theta
\right) \left( \lambda \gamma ^{t}\theta \right) \left( \theta \gamma
_{r}\psi _{g_1}\right) \left( \xi ^{2}\gamma _{t}\theta \right) \rangle \\
&&+\frac{\pi i\alpha ^{\prime }}{18}F_{m_3f_3}^{3}\langle \left( \lambda
\gamma ^{g_1}\gamma ^{m_3f_3}\theta \right) \left( \lambda \gamma ^{r}\theta
\right) \left( \lambda \gamma ^{t}\theta \right) \left( \theta \gamma
_{r}\psi _{g_1}\right) \left( \xi ^{2}\gamma _{t}\theta \right) \rangle \\
&&+\frac{\pi i\alpha ^{\prime }}{32}F_{m_2f_2}^{2}\langle \left( \lambda
\gamma _{r}\xi ^{3}\right) \left( \psi _{g_1}\gamma ^{r}\theta \right) \left(
\lambda \gamma ^{g_1}\theta \right) \left( \lambda \gamma _{p}\theta \right)
\left( \theta \gamma ^{m_2f_2p}\theta \right) \rangle \\
&&+\frac{\pi i\alpha ^{\prime }}{48}F_{m_2f_2}^{2}\langle \left( \lambda
\gamma ^{g_1}\xi ^{3}\right) \left( \psi _{g_1}\gamma _{r}\theta \right)
\left( \lambda \gamma ^{r}\theta \right) \left( \lambda \gamma _{p}\theta
\right) \left( \theta \gamma ^{m_2f_2p}\theta \right) \rangle \\
&&+\frac{\pi i\alpha ^{\prime }}{48}a_{f_2}^{2}\langle \left( \lambda \gamma
_{p}\xi ^{3}\right) \left( \partial _{m_1}\psi _{g_1}\gamma _{q}\theta \right)
\left( \lambda \gamma ^{g_1}\theta \right) \left( \lambda \gamma ^{f_2}\theta
\right) \left( \theta \gamma ^{p}\gamma ^{m_1q}\theta \right) \rangle \\
&&+\frac{\pi i\alpha ^{\prime }}{24}a_{f_2}^{2}\langle \left( \lambda \gamma
_{s}\xi ^{3}\right) \left( \partial _{m_1}\psi _{g_1}\gamma _{r}\theta \right)
\left( \lambda \gamma ^{r}\theta \right) \left( \lambda \gamma ^{f_2}\theta
\right) \left( \theta \gamma ^{s}\gamma ^{m_1g_1}\theta \right) \rangle \\
&&+\frac{\pi i\alpha ^{\prime }}{32}a_{f_2}^{2}\langle \left( \lambda \gamma
_{s}\xi ^{3}\right) \left( \partial _{m_1}\psi _{g_1}\gamma ^{s}\theta \right)
\left( \lambda \gamma _{t}\theta \right) \left( \lambda \gamma ^{f_2}\theta
\right) \left( \theta \gamma ^{m_1g_1t}\theta \right) \rangle \\
&&+\frac{\pi i\alpha ^{\prime }}{120}a_{f_2}^{2}\langle \left( \lambda \gamma
^{m_1}\xi ^{3}\right) \left( \partial _{m_1}\psi _{g_1}\gamma _{t}\theta
\right) \left( \lambda \gamma _{r}\theta \right) \left( \lambda \gamma
^{f_2}\theta \right) \left( \theta \gamma ^{rg1t}\theta \right) \rangle
\end{eqnarray*}
all the terms above are quite similar to the one graviton and two photinos
computation. Following the same steps and using the fact that the gravitino
is gamma traceless extensively we obtain

\[
\mathcal{A}_{1}=\pi i\alpha ^{\prime }\left( \frac{1}{3840}a_{f_2}\psi
_{g_1}\gamma ^{f_2}\partial ^{g_1}\xi _{3}-\frac{1}{7680}a_{f_2}\psi _{g_1}\gamma
^{f_2g_1m_3}\partial _{m_3}\xi _{3}\right) .
\]

Using now the identity (\ref{A5}), we obtain

\begin{eqnarray*}
\mathcal{A}_{1} &=&\pi i\alpha ^{\prime }\left( \frac{1}{3840}a_{f_2}\psi
_{g_1}\gamma ^{f_2}\partial ^{g_1}\xi _{3}+\frac{1}{7680}a_{f_2}\psi _{g_1}\gamma
^{f_2}\partial ^{g_1}\xi _{3}\right) \\
&=&\frac{3\pi i\alpha ^{\prime }}{7680}a_{f_2}\psi _{g_1}\gamma ^{f_2}\partial
^{g_1}\xi _{3}.
\end{eqnarray*}

Using the same argument for the other ten terms, we obtain

\[
\mathcal{A}_{2}=\frac{\pi i\alpha ^{\prime }}{2160}a_{f_2}\psi _{g_1}\gamma
^{f_2}\partial ^{g_1}\xi _{3} ,
\]

\begin{eqnarray*}
\mathcal{A}_{3} &=&\pi i\alpha ^{\prime }\left( -\frac{1}{5760}%
F_{m_3f_3}^{3}\psi ^{m_3}\gamma ^{f_3}\xi _{2}+\frac{1}{5760}F_{m_3f_3}^{3}\psi
^{f_3}\gamma ^{m_3}\xi _{2}+\frac{1}{5760}F_{m_3f_3}^{3}\psi _{g_1}\gamma
^{f_3g_1m_3}\xi _{2}\right) \\
&=&-\frac{\pi i\alpha ^{\prime }}{1440}F_{m_3f_3}^{3}\psi ^{m_3}\gamma ^{f_3}\xi
_{2},
\end{eqnarray*}

\begin{eqnarray*}
\mathcal{A}_{4} &=&-\frac{\pi i\alpha ^{\prime }}{2880}F_{m_3f_3}^{3}\psi
^{m_3}\gamma ^{f_3}\xi _{2}+\frac{\pi i\alpha ^{\prime }}{2880}%
F_{m_3f_3}^{3}\psi ^{f_3}\gamma ^{m_3}\xi _{2} \\
&=&-\frac{\pi i\alpha ^{\prime }}{1440}F_{m_3f_3}^{3}\psi ^{m_3}\gamma ^{f_3}\xi
_{2},
\end{eqnarray*}

\begin{eqnarray*}
\mathcal{A}_{5} &=&\pi i\alpha ^{\prime }\left( \frac{1}{5760}%
F_{m_2f_2}^{2}\xi ^{3}\gamma ^{f_2}\psi ^{m_2}-\frac{1}{5760}F_{m_2f_2}^{2}\xi
^{3}\gamma ^{m_2}\psi ^{f_2}-\frac{1}{46080}F_{m_2f_2}^{2}\xi ^{3}\gamma
^{f_2g_1m_2}\psi _{g_1}\right) \\
&=&\pi i\alpha ^{\prime }\left( \frac{1}{2880}F_{m_2f_2}^{2}\xi ^{3}\gamma
^{f_2}\psi ^{m_2}-\frac{1}{23040}F_{m_2f_2}^{2}\xi ^{3}\gamma ^{f_2}\psi
^{m_2}\right) =\frac{7\pi i\alpha ^{\prime }}{23040}F_{m_2f_2}^{2}\xi
^{3}\gamma ^{f_2}\psi ^{m_2},
\end{eqnarray*}

\begin{eqnarray*}
\mathcal{A}_{6} &=&\pi i\alpha ^{\prime }\left( \frac{1}{17280}%
F_{m_2f_2}^{2}\xi ^{3}\gamma ^{f_2}\psi ^{m_2}-\frac{1}{17280}F_{m_2f_2}^{2}\xi
^{3}\gamma ^{m_2}\psi ^{f_2}+\frac{1}{17280}F_{m_2f_2}^{2}\xi ^{3}\gamma
^{f_2g_1m_2}\psi _{g_1}\right) \\
&=&\pi i\alpha ^{\prime }\left( \frac{1}{8640}F_{m_2f_2}^{2}\xi ^{3}\gamma
^{f_2}\psi ^{m_2}+\frac{1}{8640}F_{m_2f_2}^{2}\xi ^{3}\gamma ^{f_2}\psi
^{m_2}\right) =\frac{\pi i\alpha ^{\prime }}{4320}F_{m_2f_2}^{2}\xi ^{3}\gamma
^{f_2}\psi ^{m_2},
\end{eqnarray*}

\begin{eqnarray*}
\mathcal{A}_{7} &=&\pi i\alpha ^{\prime }\left( \frac{1}{34560}a_{m_1}^{2}\xi
^{3}\gamma ^{g_1}\partial ^{m_1}\psi _{g_1}-\frac{1}{138240}a_{2}^{f_2}\xi
^{3}\gamma _{f_2g_1m_1}\partial ^{m_1}\psi ^{g_1}\right) \\
&=&0,
\end{eqnarray*}

\[
\mathcal{A}_{8}=-\frac{\pi i\alpha ^{\prime }}{17280}a_{f_2}^{2}\xi
^{3}\gamma ^{f_2g_1m_1}\partial _{m_1}\psi _{g_1}=0 ,
\]

\[
\mathcal{A}_{9}=-\frac{\pi i\alpha ^{\prime }}{46080}a_{f_2}^{2}\xi
^{3}\gamma ^{f_2g_1m_1}\partial _{m_1}\psi _{g_1}=0
\]
and

\[
\mathcal{A}_{10}=\pi i\alpha ^{\prime }\left( \frac{1}{1440}a_{f_2}^{2}\xi
^{3}\gamma ^{f_2g_1m_1}\partial _{m_1}\psi _{g_1}\right) =0 .
\]

We can note that the last four terms give null results because, as in the
graviton case, there is no way to contract a kinetic term of the gravitino
with a photon and a photino that gives a non null result. Adding all
results, we obtain

\begin{eqnarray*}
\mathcal{A} &=&\pi i\alpha ^{\prime }(-\frac{3}{7680}F_{g_1f_2}^{2}\psi
_{g_1}\gamma ^{f_2}\xi _{3}-\frac{1}{2160}F_{g_1f_2}^{2}\psi _{g_1}\gamma
^{f_2}\xi _{3} \\
&&-\frac{1}{1440}F_{m_3f_3}^{3}\psi ^{m_3}\gamma ^{f_3}\xi _{2}-\frac{1}{1440}%
F_{m_3f_3}^{3}\psi ^{m_3}\gamma ^{f_3}\xi _{2} \\
&&-\frac{7}{23040}F_{m_2f_2}^{2}\psi ^{m_2}\gamma ^{f_2}\xi ^{3}-\frac{1}{4320}%
F_{m_2f_2}^{2}\psi ^{m_2}\gamma ^{f_2}\xi ^{3}) \\
&=&-\frac{\pi i\alpha ^{\prime }}{720}(F_{g_1f_2}^{2}\psi ^{g_1}\gamma ^{f_2}\xi
_{3}+F_{m_3f_3}^{3}\psi ^{m_3}\gamma ^{f_3}\xi _{2})
\end{eqnarray*}

\end{document}